\documentclass[preprint,authoryear,12pt]{elsarticle} 

\usepackage{lineno}
\modulolinenumbers[1]
\usepackage{fullpage}
\usepackage{xcolor}
\usepackage{amsmath}
\usepackage[normalem]{ulem}
\usepackage{amssymb}
\usepackage{mathrsfs}
\usepackage{newfloat}
\usepackage{graphicx}
\usepackage{stmaryrd}
\usepackage{subfig}
\usepackage{multirow}
\usepackage{stackengine}
\usepackage{caption}
\usepackage{tikz}
\usepackage{multicol}
\usetikzlibrary{arrows,positioning,calc}
\usepackage{pstool}
\usepackage{textcomp}
\usepackage[hidelinks]{hyperref}
\hypersetup{colorlinks,bookmarksopen,linkcolor=blue,citecolor=blue,urlcolor=blue}
\usepackage{float}
\usepackage{makecell}

\definecolor{myred}{RGB}{222,45,38}
\definecolor{myblue}{RGB}{0,115,189}
\definecolor{mygreen}{RGB}{49,156,54}

\newcommand{\bs}[1]{{\boldsymbol{#1}}}
\newcommand{\tensor}[1]{{\boldsymbol{#1}}}
\newcommand{\tensorfour}[1]{{}^4{\bs{#1}}}
\newcommand{\mtrx}[1]{{\underline{#1}}}
\newcommand{\column}[1]{{\underline{#1}}}

\graphicspath{{figs/}}

\journal{Int.~J. Solids Struct.}

\begin{document}

\begin{frontmatter}
	\title{Role of inter-fibre bonds and their influence on sheet scale behaviour of paper fibre networks\tnoteref{titlefoot}}	

	\author[TUe,ULB,M2i]{P. Samantray}
	\ead{priyam.samantray@gmail.com}

	\author[TUe]{R.H.J.~Peerlings}
	\ead{R.H.J.Peerlings@tue.nl}

	\author[ULB]{T.J.~Massart}
	\ead{Thierry.J.Massart@ulb.be}

	\author[TUe]{O.~Roko\v{s}\corref{correspondingauthor}}
	\ead{O.Rokos@tue.nl}

	\author[TUe]{M.G.D.~Geers}
	\ead{M.G.D.Geers@tue.nl}

	\address[TUe]{Mechanics of Materials, Department of Mechanical Engineering, Eindhoven University of Technology, P.O.~Box~513, 5600~MB~Eindhoven, The~Netherlands}
	\cortext[correspondingauthor]{Corresponding author.}

	\address[ULB]{Building, Architecture and Town Planning, Universit\'{e} Libre de Bruxelles, 50 Avenue F.D. Roosevelt, CP 194/2, B--1050 Brussels, Belgium}

	\address[M2i]{Materials innovation institute (M2i), P.O. Box 5008, 2600 GA Delft, The Netherlands}

	\tnotetext[titlefoot]{The post-print version of this article is published in \emph{Int.~J. Solids Struct.}, \href{https://doi.org/10.1016/j.ijsolstr.2022.111990}{10.1016/j.ijsolstr.2022.111990}}
	%
	%
	\begin{abstract}
		In fibrous paper materials, an exposure to a variation in moisture content causes changes in the geometrical and mechanical properties. Such changes are strongly affected by the inter-fibre bonds, which are responsible for the transfer of the hygro-mechanical response from one fibre to its neighbours in the network, resulting in sheet-scale deformation. Most models developed in literature assume perfect bonding between fibres. In the 3D reality, there is some flexibility in the bond region, even for the perfectly bonded fibres, because of the possibility of deformation gradients through the fibre thickness. In earlier 2D idealizations, perfectly bonded fibres were assumed, implying full kinematic constraint through the entire thickness of the sheet. The purpose of the present study is to assess the effect of this assumption. Using a homogenization approach, a random network of fibres is generated with different coverages and modelled using finite elements. In order to understand the role of bonding between fibres on the hygro-expansive behaviour of a network, a bond model is developed. In this model, the fibres are modelled using 2D regular bulk finite elements and the bonds are represented by interfacial elements of finite stiffness, which are introduced between each pair of fibres bonded in the network. These embedded interfacial elements form a connection between two respective fibres, allowing relative displacements between their mid-planes. The hygro-elastic response of networks obtained with this bond model is investigated by varying the bond stiffness and the network coverage under the application of mechanical loading and changes in moisture content. Furthermore, the bond model is used to analyse the influence of inter-fibre bonds on the anisotropic response of the paper fibre network.
	\end{abstract}


\end{frontmatter}

%
%
\section{Introduction}
\begin{figure}
	\centering
	\subfloat[a micrograph of paper fibres]{\includegraphics[height=50mm]{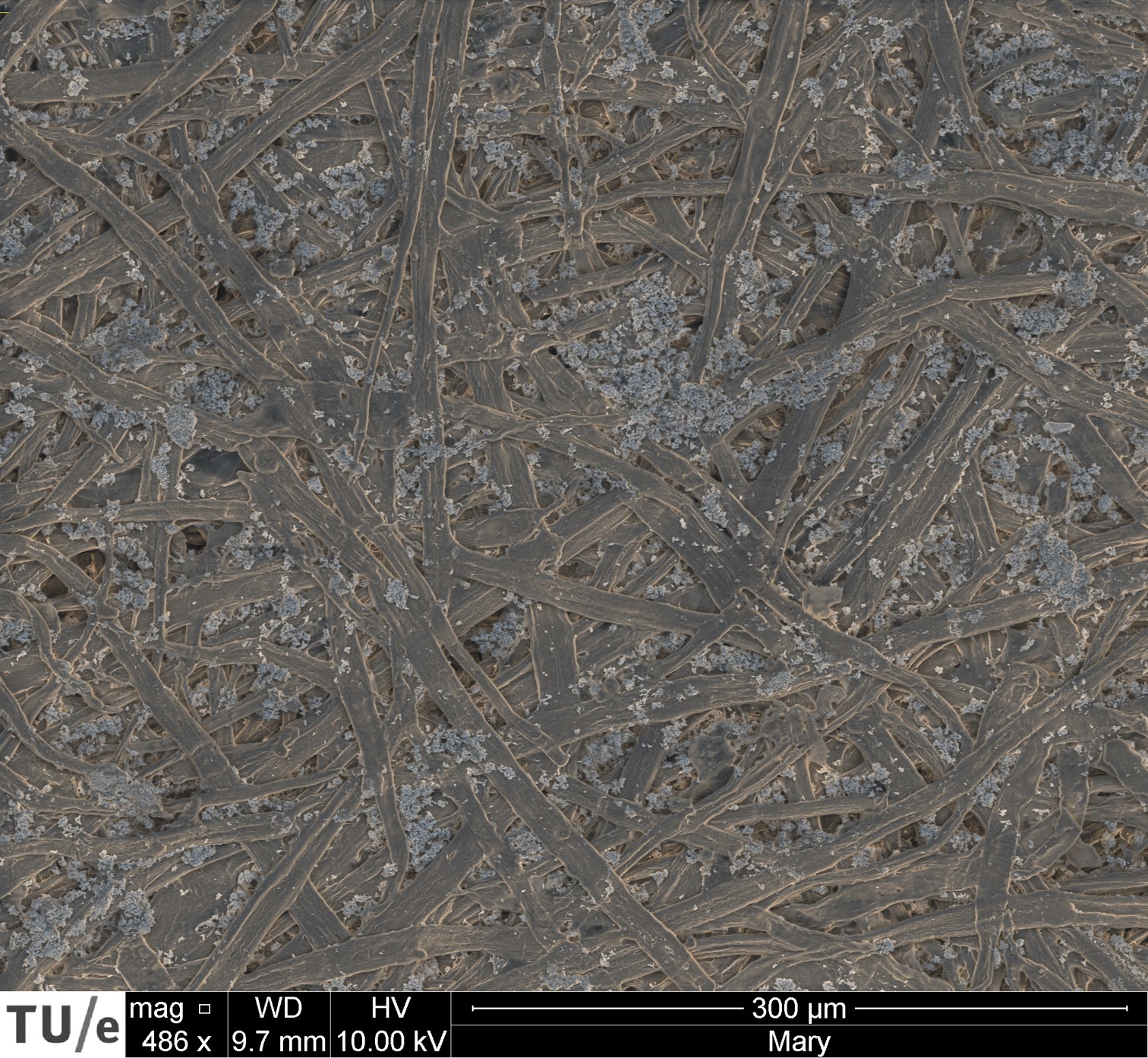}\label{edgi14}}
	\hspace{1em}
	\subfloat[a random paper fibre network]{\includegraphics[height=50mm]{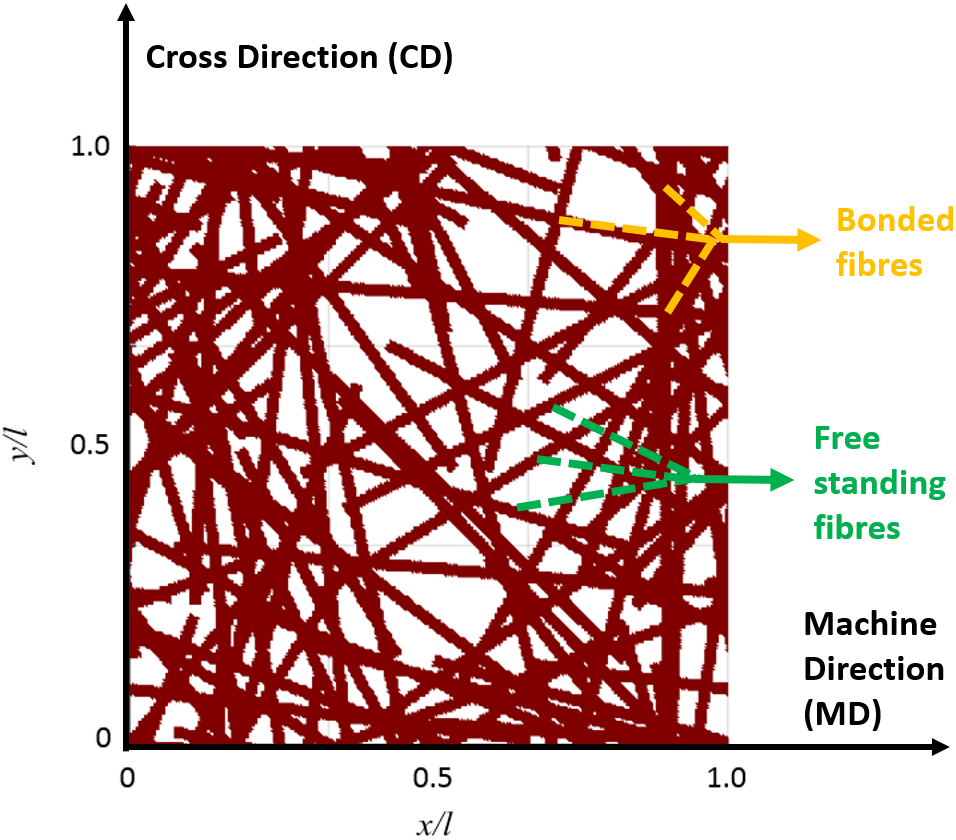}\label{edgi24}}
	\caption{(a)~A micrograph of paper showing random orientation of individual fibres, and~(b) an artificially generated network of random paper fibres subjected to hygroscopic strain. The preferential directions are the Machine Direction~(MD), oriented along~$\vec{e}_x$, and the Cross Direction~(CD), oriented along~$\vec{e}_y$.}
\end{figure}

\begin{figure}
	\centering
	\hspace{7em}
	\subfloat[wall layer structure of  a wood fibre]{\includegraphics[height=45mm]{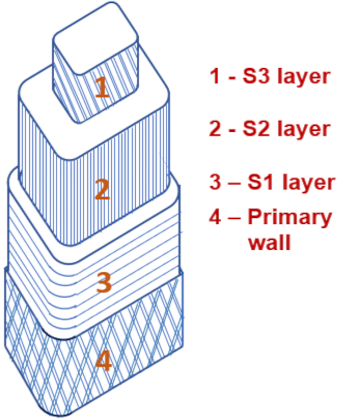}\label{layer1}}
	\hspace{3em}
	\subfloat[ultrastructural organization of  cellulose within the wood cell wall]{\includegraphics[height=45mm]{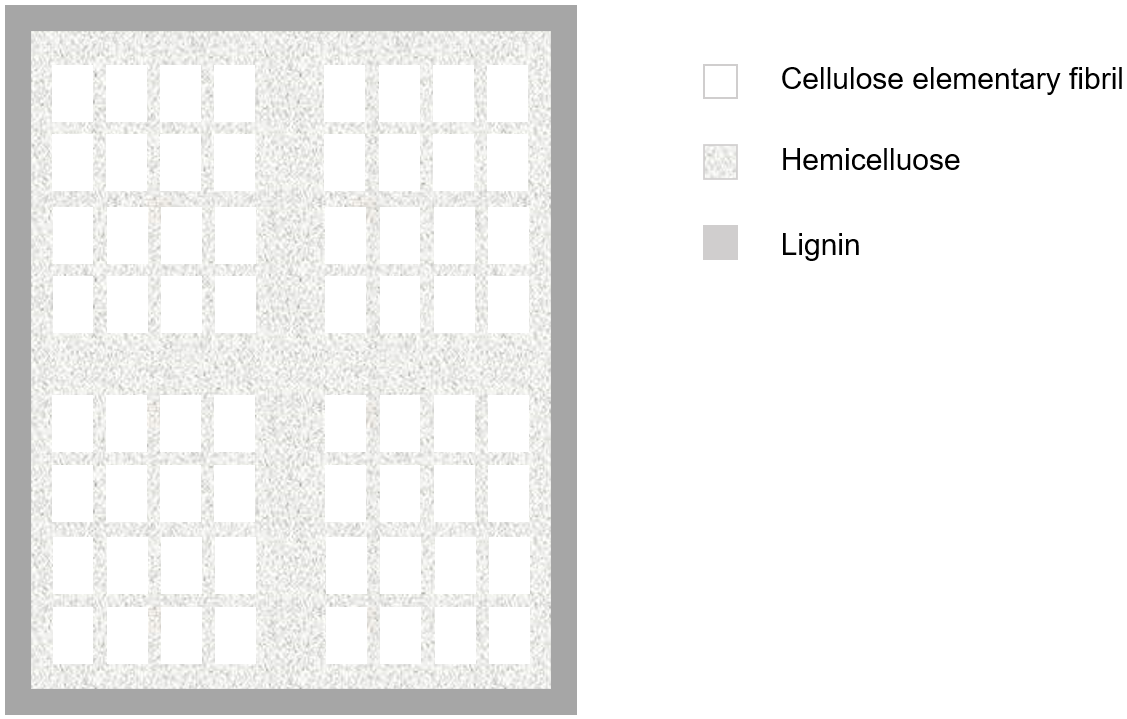}\label{ultra}}
	\caption{(a)~A schematic diagram of different cell wall layers within a wood fibre, adapted from~\citep{Brandstorm,Gilani}. (b)~Organisation of cellulose polymers at the ultrastructural level within a wood cell wall, adapted from~\citep{Neagu}.}
	\label{fig:fibre_structure}
\end{figure}

Fibrous materials like paper consist of natural fibres that are bonded in overlapping regions, as shown in Fig.~\ref{edgi14}. In paper in particular, individual fibres are hollow layered structures made of wood pulp, which consist of a primary cell wall on the outer side and a secondary cell wall~S. The secondary cell wall further consists of an outer layer~S1, middle layer~S2, and an inner layer~S3, as indicated in Fig.~\ref{layer1}. Individual cell walls are furthermore composed of three types of polymers (in particular cellulose, hemicellulose, and lignin), having the ultrastructural composition shown in Fig.~\ref{ultra}. The distribution of the polymers varies across the primary and secondary walls of the fibre, where the secondary wall layer contains higher amounts of cellulose and hemicellulose as compared to the primary layer, whereas lignin content is nearly the same among all the secondary layers, cf.~\citep{Neagu}.

During the manufacturing of paper, the fibres are preferentially oriented in the machine direction, which entails anisotropy in the mechanical behaviour of the paper network~\citep{Larsson}. The principal directions of the paper material are denoted as the machine direction (MD) and cross direction (CD), as shown in the random fibre network in Fig.~\ref{edgi24} \citep{Bosco1}. Here, individual fibres are interconnected to each other in overlapping regions, called bonds, complemented with freely standing parts in between bonds. In a paper fibre network, the fibres in inter-fibre bonds are connected together by hydrogen bonding and van der Waal's forces~\citep{Henrikkson}. The anisotropic behaviour, originating at the scale of individual fibres, is transferred from one region in the network to another through these inter-fibre bonds. The coefficient of transverse hygro-expansion of individual fibres is nearly $ 20 $ times the corresponding value in the longitudinal direction~\citep{Niskanen1}. This is because at the fibril scale, the polymers (i.e., hemicellulose, cellulose, and lignin) exhibit hygro-expansion that varies depending on multiple factors, which in combination with the complex fibre microstructure (recall Fig.~\ref{fig:fibre_structure}) results in high anisotropy. For example, the hygro-expansion of the hemicellulose has been found to be highest among all three polymers due to high solubility of its side groups~\citep{hansen}, whereas cellulose (in crystalline form) has negligible hygro-expansion, as shown by several studies~\citep{Lindner, Marklund, Ningling, Cave}. The water uptake by cellulose decreases with an increase in crystallinity below 75\% relative humidity~\citep{Mihranyan}, although the wood fibre has 60--70\% cellulose chains in crystalline form and only the rest is in water-absorbing amorphous form and hence contributing to hygro-expansion~\citep{Alince}. Apart from crystallinity, the water absorption in cellulose depends on the temperature at which the fibres are processed, on the solvents around cellulose, the surrounding cellulose molecules, and the processing of the fibres~\citep{Fahlen,Zeta,Berthold}. For lignin, the moisture content varies depending on the type, i.e., 3--5\% for dioxane lignin and 10--15\% for Klason and periodate lignin. Other parameters affecting hygro-expansivity of lignin include softening temperature~\citep{eriksson} and microfibril angle~\citep{Ningling}.


The degree of bonding between fibres and the number of fibres in the network have a crucial influence on the hygro-mechanical response of the network. Earlier studies demonstrated that the effective stiffness of the network is mainly dependent on the density of the network, fibre orientations, fibre properties, and of the inter-fibre connectivity resulting from the bonds~\citep{Mao1,Karakoc,Heyden,bergstrom,Cox}. It was furthermore revealed that when subjected to moisture changes, the effective hygro-expansivity of the paper network is governed mainly by the bonded regions of the network~\citep{Bosco3,Sam1,Uesaka1,Uesaka2,Vainio,Niskanen1,Larsson,Erkilla,Nordman,Salmen,Schulgasser}. In another study~\citep{mark}, the properties of the bonded fibres were computed from load-displacement measurements. The hygro-mechanical properties of the paper, which are highly dependent on the inter-fibre connectivity, thus hold significance in understanding the dimensional stability of paper products like packaging materials, printed papers, corrugated boxes, tissue papers, or paper boards.  Instabilities appearing as curls or waviness at the paper sheet scale are moreover significantly dependent on bond stiffness. In this study, the main objective is therefore to examine the effect of the bond stiffness on the sheet-scale response of paper network, and to critically review the assumption often made in earlier works (i.e., full kinematic tying in the inter-fibre bonds).

Several earlier works consider bonds between the fibres to be rigid, for both 2D and 3D configurations~\citep{Bronkhorst,Mao1,Bosco2,Starzewski,Lee,Stahl}. A limitation of such a bond description is its inability to describe the influence of inter-fibre connectivity on the sheet scale deformations, as well as the extent to which certain fibre network properties (e.g., bond area or bonding stiffness) affect the macroscale response that can be modified in the manufacturing process. Some other works, on the other hand, described the bonds as flexible and not rigid, using for instance springs~\citep{Heyden}. The fibres were partially bonded in regions of overlap~\citep{Koh}, and were modelled based on a non-linear contact law with bond failure~\citep{Kulachenko}, or as two-node line elements~\citep{Liu}. Studies by~\cite{Rama} and~\cite{Perkins} consider the shear stresses and inelastic behaviour to be present in the bonds, hence assuming their deformability. However, most of the literature that incorporates the deformation in bonds relates only to the response of networks under mechanical loading. Analysis and modelling of the hygro-mechanical behaviour and the anisotropic response of the network due to non-rigid bonds are not available, which especially in 3D entail costly computations that prevent assessing parametric variations of the bond effect. Even though for hygro-mechanics only perfect bonding is relevant, a perfect bonding in 3D reality is not equivalent to full compatibility in 2D. There is therefore a need to understand the hygro-mechanical response of the networks due to varying bonding properties (stiffness) connecting the fibres, along with its influence on the anisotropy of the network using a computationally efficient 2D modelling framework.

In this contribution, such issues are addressed by describing the inter-fibre connectivity through an appropriate 2D bond model, starting with a periodic unit cell of randomly generated rectangular strips~\citep{Sam2} that represent ribbon-shaped paper fibres. Each fibre in the unit cell network is discretized with standard triangular finite elements. At the regions of inter-fibre bonds, additional triangular elastic interfacial elements with a suitable stiffness, connecting two successive fibres (in the thickness direction), are inserted in the model. These finite stiffness elements render the bonds to be no longer rigid and the connected fibres can therefore have relative displacements through induced elasticity. Note that, although possible, no dissipation within the interfacial layer (plasticity nor damage) is assumed within the inserted layer, and hence modelling of degradation of the bond is not adopted in this work. With this model, the macroscopic behaviour of the paper network with different values of the bond stiffness can be computed when subjected to an external mechanical load or a uniform moisture variation. This provides clear insights in the sheet level response using a 2D network model with moderate computational efforts only, while incorporating key elements of the full 3D reality. The influence of the connectivity between fibres at the bonds on the anisotropy of the paper network at the sheet scale can furthermore be assessed.

The contents of this manuscript is organized as follows. In Section~\ref{sec:constitutive}, the hygro-mechanical model for the fibre and the network is described. This includes the generation of random networks, the bond model, and the adopted numerical discretization together with considered stacking of fibres in the network. Obtained results, illustrating the influence of the bond stiffness at the sheet level response of the network along with its anisotropy, are detailed in Section~\ref{sec:results}. The conclusions are finally drawn in Section~\ref{sec:conclusions}.

Throughout this contribution, the following notation is used for operations on Cartesian tensors and for tensor products. Scalars, vectors, and tensors are denoted as~$a$, $\vec{{a}}$, and $\tensor{A}$. For tensor and vector operations, the following equivalent notations are used, with Einstein's summation convention on repeating indices: $\tensor{A}:\tensor{B} = A_{ij}B_{ji}$, where~$i = x, y, z$, for the global reference system, and~$i = l,t,z$, for the fibre local reference system. The Voigt notation employed to represent tensors and tensor operations in a matrix format uses~$\column{a}$ and~$\mtrx{A}$ to denote a column matrix and a matrix of scalars; matrix multiplication then reads $\mtrx{A}\,\column{b} = A_{ij}b_j$.
%
%
\section{Hygro-mechanical constitutive model and numerical discretization}
\label{sec:constitutive}
%
%
\subsection{Fibre model}
Consider a 2D configuration with plane stress assumption in the \textit{z}-direction (with \textit{z} normal to the paper sheet). This assumption is adopted because a fully-complex 3D model may easily become computationally too expensive, especially for parametric studies. The 3D reality is hence simplified by considering only the mid-planes of the fibres, see Fig.~\ref{new3d2d44} where a pair of bonded fibres is depicted. Such assumption is adopted because the main objective of this work is to seek information about the role of stiffness in the bond on the hygro-expansion and anisotropy of the paper networks, with affordable computational costs. Since computational efforts are limiting in full 3D simulations, the herein proposed model is formulated in two dimensions while accounting for the key mechanisms of a full three-dimensional reality through bond elasticity. Earlier-mentioned instabilities, such as curls or waviness, could be incorporated within the proposed model by introducing shell elements instead of plane stress elements. Under such assumptions, the hygroscopic strain tensor ${}^h\tensor{\varepsilon}^f$ and the stress tensor in a fibre~$ f $ are given by the constitutive relations
\begin{align}
	{}^h\tensor{\varepsilon}^f & = \tensor{\beta}^f\Delta\chi, \label{epsh4}                                   \\
	\tensor{\sigma}^f          & = \tensorfour{C}^f : ( \tensor{\varepsilon}^f - {}^h\tensor{\varepsilon}^f ),
\end{align}
where $\tensor{\sigma}^f$, $\tensorfour{C}^f$, $\tensor{\varepsilon}^f$, and $\tensor\beta^f$ are the stress tensor, the elastic stiffness tensor, the strain tensor, and the hygroscopic expansion tensor of the fibre. The variable~$ \Delta\chi = \chi - \chi_0$ then denotes change in the moisture content relative to the reference value~$\chi_0$, where the moisture content $\chi$ is defined as the ratio of the mass of water in the paper relative to the total mass of the paper. For this study, applications within digital printing are mostly assumed, for which the moisture in paper varies from~2\% to 10\% based on~\citep{bullet}, depending also on the atmospheric conditions such as relative humidity, temperature, past usage, and type of paper. The elastic strain tensor of the considered fibre then reads~$ {}^e\tensor{\varepsilon}^f =  \tensor{\varepsilon}^f - {}^h\tensor{\varepsilon}^f $. In the Voigt matrix notation, the stiffness and hygro-expansion coefficients are represented as
\begin{equation}
	\mtrx{C}^f=
	\begin{pmatrix}
		\frac{E_l}{(1-\nu_{lt}\nu_{tl})}         & \frac{E_l\nu_{tl}}{(1-\nu_{lt}\nu_{tl})} & 0      \\
		\frac{E_t\nu_{lt}}{(1-\nu_{lt}\nu_{tl})} &
		\frac{E_t}{(1-\nu_{lt}\nu_{tl})}         & 0                                                 \\
		0                                        & 0                                        & G_{lt}
	\end{pmatrix},
	\quad
	\column{\beta}^f =
	\begin{pmatrix}
		\beta_l \\
		\beta_t \\
		0
	\end{pmatrix},
	\label{mat4}
\end{equation}
where~$E_l$ and~$E_t$ are the elastic moduli in the longitudinal and transverse direction with respect to the fibre material axes, $\beta_l$ and~$\beta_t$ represent the coefficients of hygroscopic expansion in the longitudinal and transverse directions associated to the fibre axis, $\nu_{lt}$ and~$\nu_{tl}$ are the Poisson's ratios, and~$G_{lt}$ is the in-plane shear modulus. All material properties are assumed to be independent of the moisture content $\chi$ within this study, although in reality the dependence on moisture content is significant. In particular, the Young's modulus of a dry fibre is~$2.5$ times the Young's modulus of a fully wet fibre, as reported by~\cite{Jentzen}. Even though such a simplifying assumption may have consequences on a quantitative behaviour of the paper sheet, it still has relevance in providing the desired qualitative knowledge about the sheet scale behaviour of paper under varying stiffness in the bonds. The constitutive relationships in Eqs.~\eqref{epsh4}--\eqref{mat4} are expressed in the local reference system ($l,t,z$), attached to the fibre as sketched in Fig.~\ref{axi4}, for a fibre~$f$ oriented at an angle $\theta^{(f)}$ with respect to MD. This relationship can be transformed to the global reference system ($x,y,z$) yielding the global elastic constitutive tensor and hygroscopic coefficient tensor~\citep{Roylance}. All the fibres are assumed to have a uniform thickness.
%
%
\subsection{Bond model}
\begin{figure}
	\centering
	\subfloat[global $ ( x, y, z ) $ and local $ ( l, t, z ) $ coordinate frames]{\includegraphics[width=0.35\linewidth]{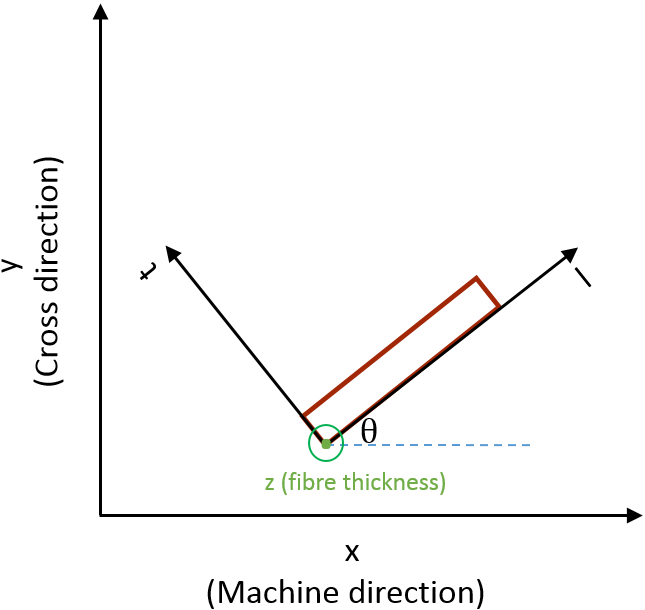}\label{axi4}}
	\hspace{2em}
	\subfloat[3D versus 2D representation of a pair of fibres]{\includegraphics[width=0.55\linewidth]{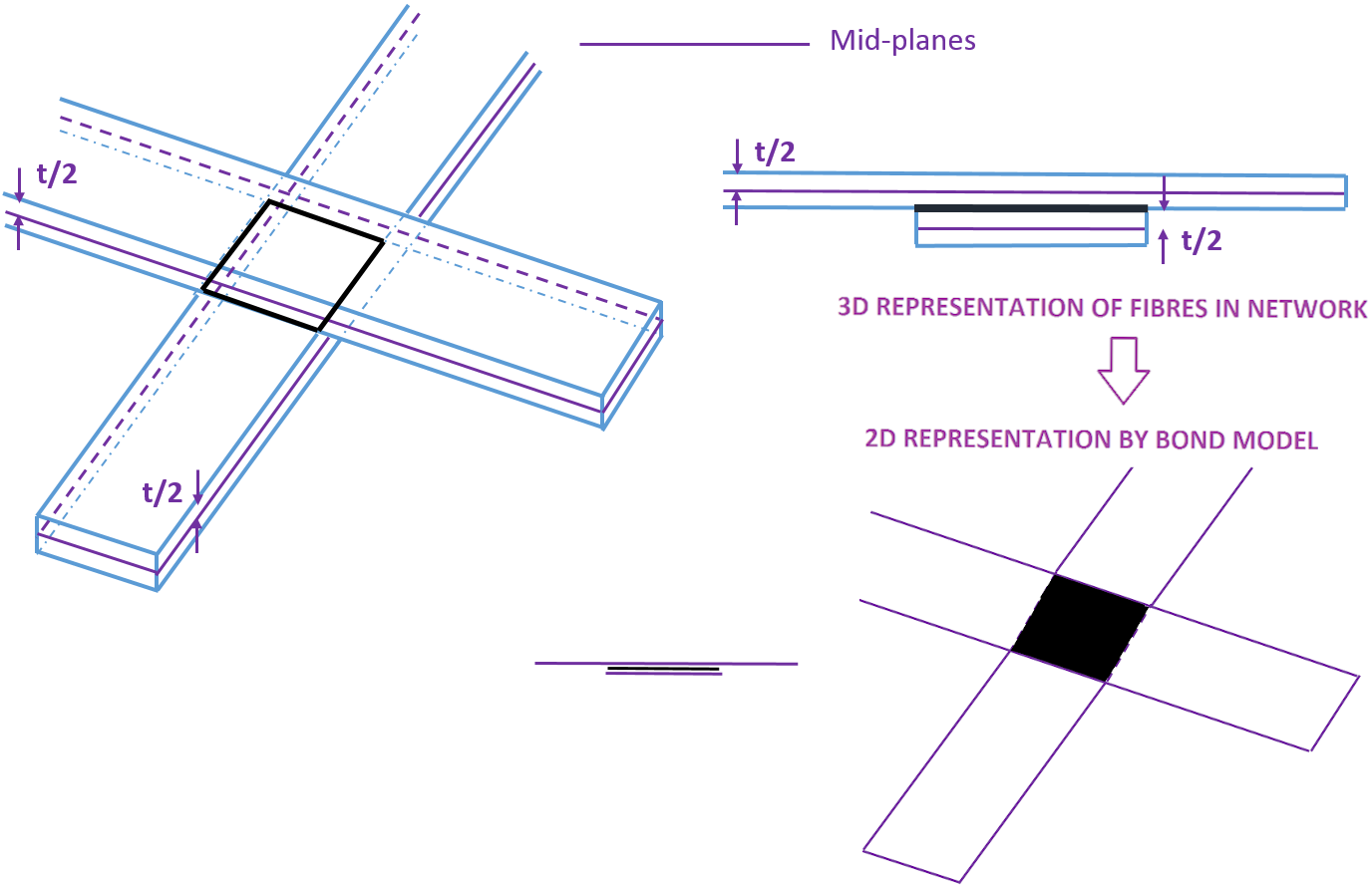}\label{new3d2d44}}
	\caption{(a)~The global and local coordinate systems considered for the description of fibre's geometry; the depicted fibre is oriented by an angle~$\theta$ relative to the Machine Direction~(MD) along the~$ \vec{e}_x $ axis. (b)~Representation of a pair of fibres considered in 3D, and in 2D according to the proposed bond model.}
	\label{fibres}
\end{figure}

\begin{figure}
	\centering
	\subfloat[two bonded fibres in 3D]{\includegraphics[width=0.25\linewidth]{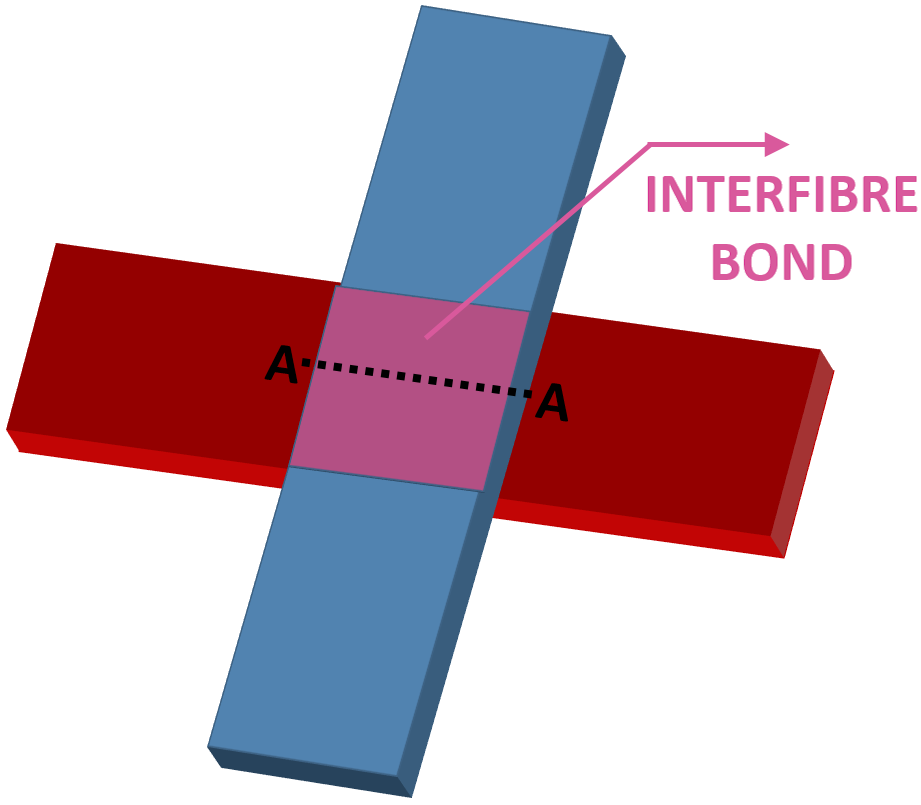}\label{3d4}}
	\hfill
	\subfloat[three types of bonds over the cross-section A--A]{\includegraphics[width=0.7\linewidth]{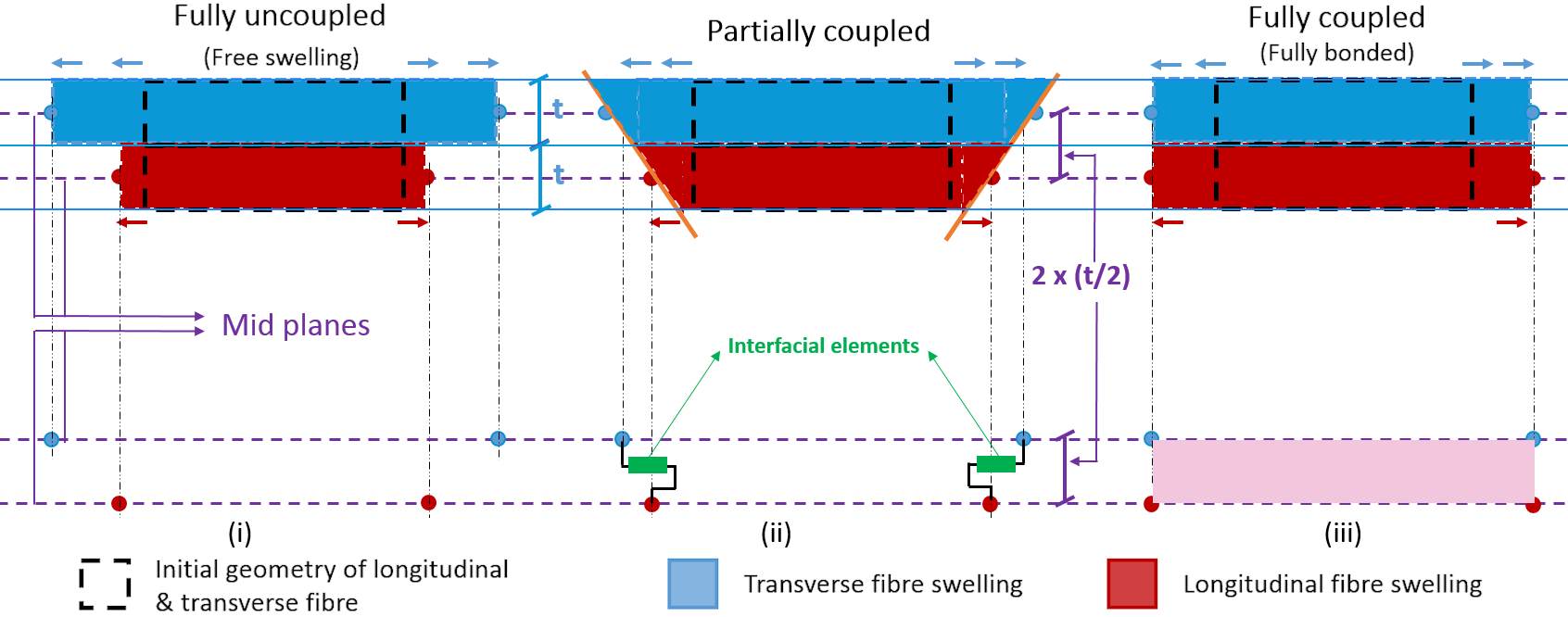}\label{mech4}}
	\caption{The mechanism of the bond model interconnecting overlapping fibres. Note that in the sketch the fibres are perpendicular to each other for clarity, although in reality they need not be perpendicular. (a)~Two bonded fibres depicted in 3D, and considered cross-section A--A. (b)~Three types of bonding of the two fibres along the A--A cross-section: fully uncoupled (no bonding; left), partially coupled (elastic bonding; middle), fully coupled (rigid bonding; right). Note that the blue and red bullets represent finite element nodes.}
	\label{ne4}
\end{figure}

In the network, regions in which the fibres overlap are called inter-fibre bonds, as shown in Figs.~\ref{new3d2d44} and~\ref{3d4}. In a realistic 3D context, the mechanics between fibres in the bond is relatively complex, and is represented by a shear modulus $G_{lt}$. The type of the bond between fibres can be represented with three interconnect models, depicted in the cross-section A--A of Fig.~\ref{ne4}. In the first, fully uncoupled case, the transverse and longitudinal regions of the fibres involved in the bond are completely detached from each other, resulting in free swelling across the mid-planes. In the fully coupled case, on the contrary, both fibres are connected by a rigid bond, entailing compatible displacements in the bond plane (and thus also in the mid-plane), as is often used in the literature~\citep{Bosco1}. In the last, partially coupled case, the fibres are constrained to some extent, but may exhibit relative displacements. This closely resembles a case in which springs interconnecting the fibres are considered to enforce kinematic coupling. The fibres are thus partially tied to each other, and do not exhibit free swelling.


To implement partial coupling, an elastic interface model is added (i.e., free of any dissipation such as plasticity, sliding, or damage of the interface), which represents the compliance contained in the layer of a thickness~$t$ between the mid-planes of the fibres. This interface thus connects the two fibres in the bond, allowing them to deform with respect to each other, and thereby relaxing the rigid constraint of the fully coupled case. The extent to which the stacked fibres may relax depends on the adopted interface stiffness. All the displacements are confined in the~$x$--$y$ plane, and hence the relative displacements are only tangential.

Let us consider a bond which is governed by the following constitutive law
\begin{align}
	\vec{\tau}    & = \frac{k_h}{t}{\Delta \vec{g}},                 \\
	\Delta\vec{g} & = \Delta u \, \vec{e}_x + \Delta v \, \vec{e}_y,
	\label{eq:constitutive_bond}
\end{align}
where~$\vec{\tau}$ is the shear stress vector acting in the plane of the interface, $k_h$ is the interfacial stiffness modulus of the bond between both fibres, having the same units as the shear modulus $G_{lt}$ of a fibre, $\Delta \vec{g}$ is the differential displacement vector, consisting of differential displacements between the fibre mid-planes in the~$\vec{e}_x$ and~$\vec{e}_{y}$ directions, denoted~$\Delta u$ and~$\Delta v$, and~$t$ is the thickness of a fibre. For the sake of convenience we will represent~$G_{lt}$ by~$G$ hereafter. When~$k_h = G$, the bond represents the physical shear between the mid-planes of both fibres, taking their thickness into account for this 2D setting. We thus expect~$k_h$ to be of the order of the shear modulus~$G$, although it may not be exactly equal to it. The hygro-mechanical response of the networks at different values of interfacial stiffness~$k_h$ is illustrated later in the results section to render a complete picture in this regard. Note that although not adopted herein, potential bond degradation such as plasticity, sliding, or damage, could readily be implemented within the constitutive law of Eq.~\eqref{eq:constitutive_bond}, by introducing proper internal variables and dissipation mechanisms.
%
%
\subsection{Network model}
\label{sec:networkModel}
\begin{figure}
	\centering
	\subfloat[random fibrous network, $c = 0.5$]{\includegraphics[height=60mm]{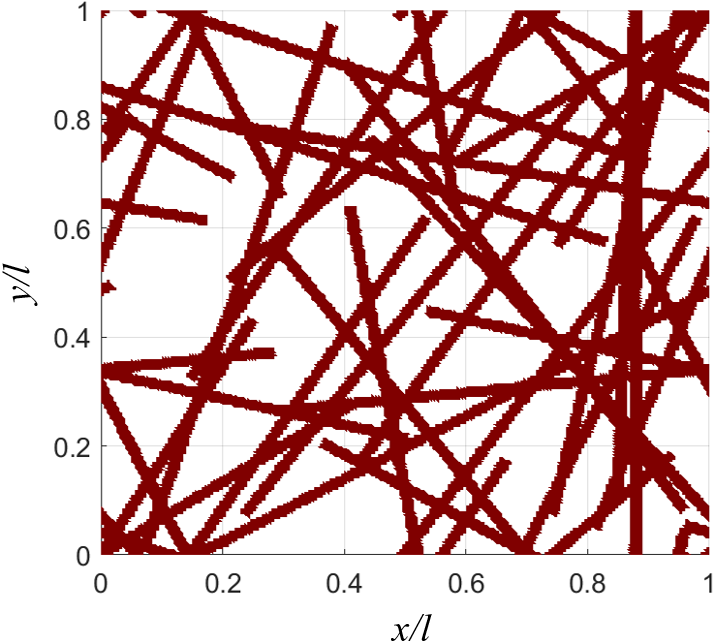}\label{ran444}}
	\hspace{1em}
	\subfloat[simple network of three fibres]{\includegraphics[height=60mm]{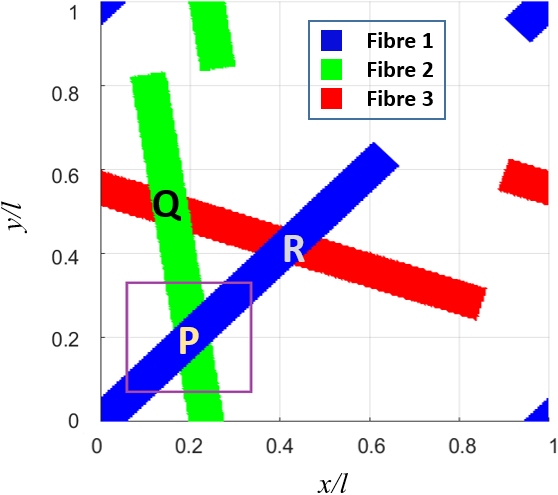}\label{stack4}}
	\caption{(a)~Periodic unit cell of a random fibrous network, having coverage~$c = 0.5$. (b)~A simple sparse fibre network, consisting of only three fibres.}
	\label{ranstack}
\end{figure}

In a 2D representation, the paper fibres are assumed to be ribbon-shaped elements. Hence, networks are represented by a set of randomly generated rectangular fibres. The fibre length is denoted~$l_f$, the fibre width~$w_f = l_f/50$, and the thickness is assumed to be~$t = l_f/150$. The positions of the centroids of the rectangular fibres are randomly generated within the domain $[x,y] \in [0,l] \times [0,l]$ of the periodic unit cell of length~$l = l_f$, as shown in Fig.~\ref{ran444}. More details on the periodicity implementation follows below in this section. During the generation of fibres, individual fibres are numbered and assumed to be stacked in the corresponding sequence, shown in Fig.~\ref{stack4}. The orientation of the fibres in the network is also random. In the particular case of a uniform orientation probability density, the corresponding network is expected to be isotropic. To characterize the number of fibres added to a given periodic unit cell, the coverage~$c$ is defined as the ratio of the area of all fibres relative to the area of the unit cell, a number that will be used later in this study. The coverage is a measure of the areal density of fibres in the network, denoted~$c$, i.e., the grammage.

\begin{figure}
	\centering
	\subfloat[Layer~1]{\includegraphics[height=56mm]{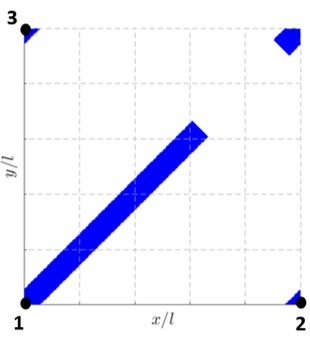}\label{fig:controlPoints}}
	\hspace{1em}
	\subfloat[Layer~2]{\includegraphics[height=56mm]{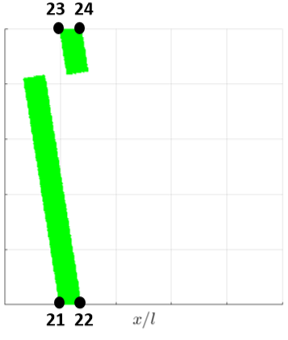}}
	\hspace{0.5em}
	\subfloat[Layer~3]{\includegraphics[height=56mm]{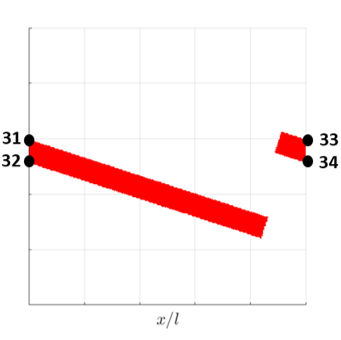}}
	\caption{Layers associated with individual fibres of the simple three-fibre network of Fig.~\ref{stack4}. Each fibre, 1, 2, and~3 (cf. Fig.~\ref{stack4}), has its own layer, 1 in~(a), 2 in~(b), and~3 in~(c), associated with it. Individual fibres are bonded in overlapping regions. RVE control points are the points~$1$, $2$, and~$3$ of the Layer~$1$ in~(a).}
	\label{period4}
\end{figure}

Each of the fibres in the network is assumed to occupy a particular layer pertaining to the fibre number, as shown in Fig.~\ref{period4}. In order to compute the hygro-mechanical response of such RVE, a finite element based method is adopted. With this method, and using a structured triangular mesh, the finite elements and nodes are created to represent only the fibres in each of the layers. They are numbered in a sequential manner with each fibre, i.e., fibre after fibre or equivalently layer by layer in the mesh. A stack of elements is thus created, representing the stack of overlapping fibres.

\begin{figure}
	\centering
	\includegraphics[height=50mm]{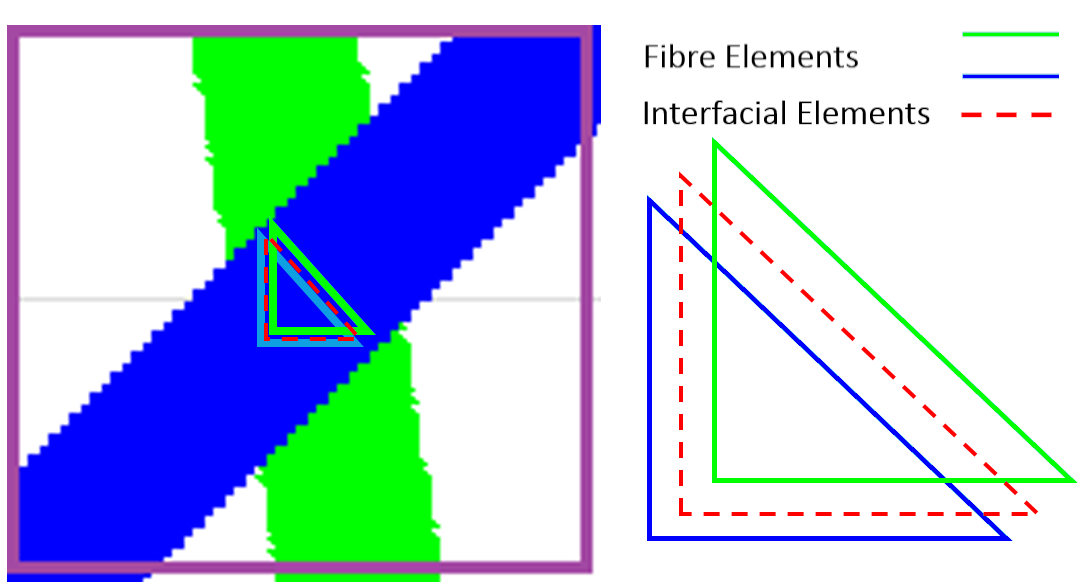}
	\caption{Triangular Interfacial elements (dashed line in red colour) and triangular standard finite elements representing the fibres (depicted as solid lines in blue and green colour) are considered at the bonds for performing the finite element simulation.}
	\label{inter44}
\end{figure}

The fibres that overlap a given fibre and are immediately below it are assumed to be bonded to it, and triangular interfacial elements are introduced accordingly (consistently with the triangular discretization of all fibres), see Fig.~\ref{inter44}. Such interfacial elements at the inter-fibre bonds are thus defined between stacked fibres at a point of the RVE, linking fibres by pairs in an ascending sequence of their layer number. For example, in Fig.~\ref{stack4}, it can be seen that fibres~$1$ and~$2$ have a bond at region P. Similarly, fibres~$1$ and~$3$ (where fibre~$1$ is in a layer not immediately below the layer of fibre~$3$) are connected through an inter-fibre bond at region R.

Each of the layers is considered periodic and modelled accordingly. The first layer contains three control nodes of the RVE, numbered~$1$, $2$, and~$3$, in Fig.~\ref{fig:controlPoints}. The boundary conditions are implemented with respect to these control nodes for each of the fibres. Nodes are assigned at the boundary of the layers, whenever a fibre crosses the boundary. For instance, periodicity implementation for layers~$2$ and~$3$ is given by
\begin{equation}
	\begin{aligned}
		\mbox{Layer 2: } & \vec{g}_{23}-\vec{g}_{21} = \vec{g}_{3}-\vec{g}_{1}, \\
		                 & \vec{g}_{24}-\vec{g}_{22} = \vec{g}_{3}-\vec{g}_{1}, \\
		\mbox{Layer 3: } & \vec{g}_{33}-\vec{g}_{31} = \vec{g}_{2}-\vec{g}_{1}, \\
		                 & \vec{g}_{34}-\vec{g}_{32} = \vec{g}_{2}-\vec{g}_{1},
	\end{aligned}
\end{equation}
where~$\vec{g}$ is the displacement vector. The fibres are thus allowed to displace relative to each other and relative to the control nodes, while the periodicity of the network is preserved. In addition to periodic boundary conditions, rigid body motions are factored out by fixing corresponding directions of the three control nodes.
%
%
\subsection{Macroscopic response}
The overall hygro-mechanical response of a network can be assessed by its effective macroscopic properties. These depend on the material and geometrical properties of the individual fibres interacting in the bonded regions. At the sheet scale, the hygro-expansive strain under stress-free conditions due to a uniform change in moisture content~$\Delta\chi$ can be defined as
\begin{equation}
	{}^h\overline{\tensor{\varepsilon}} = \overline{\tensor{\beta}} \Delta\chi,
\end{equation}
where~$\overline{\tensor{\beta}}$ is the effective hygro-expansive coefficient tensor of the paper network. The total strain~$\overline{\tensor{\varepsilon}}$ is represented by an additive decomposition into the elastic~${}^e\overline{\tensor{\varepsilon}}$ and hygro-expansive~${}^h\overline{\tensor{\varepsilon}}$ strain, i.e.,
\begin{align}
	\overline{\tensor{\varepsilon}} & = {}^e\overline{\tensor{\varepsilon}} + {}^h\overline{\tensor{\varepsilon}},                            \\
	\overline{\tensor{\sigma}}      & = {}^4\overline{\tensor{C}} : (\overline{\tensor{\varepsilon}} - \overline{\tensor{\beta}} \Delta\chi).
	\label{constit4}
\end{align}
Here, Eq.~\eqref{constit4} represents the constitutive relation at the sheet scale in which~${}^4\overline{\tensor{C}}$ represents the effective elastic stiffness tensor and~$\overline{\tensor{\sigma}}$ the macroscopic stress tensor applied to the network.
%
%
\subsection{Numerical discretization and solution}
In the periodic unit cell consisting of~$n$ fibres, the total energy~$\mathcal{E}$ associated with a unit change in moisture content is the sum of the strain energy contributions from the~$n$ fibres and the~$s$ interfacial elements over the entire volume~$V$ under a uniform moisture gradient~\citep{Chandru}, which can be represented as
\begin{align}
	\mathcal{E} & = \frac{1}{2} \int_V \sum_{i=1}^{n} (\tensor{\varepsilon}^f_i-{}^h\tensor{\varepsilon}^f_i) : \tensorfour{C}^f_i : (\tensor{\varepsilon}^f_i - {}^h\tensor{\varepsilon}^f_i) \, \mathrm{d}V + \frac{1}{2} \int_{A} \sum_{j=1}^{s} \frac{k_h}{t} \, \Delta\vec{g} \cdot \Delta\vec{g} \, \mathrm{d}A,
\end{align}
where quantities related to individual fibres in the first integral are considered non-zero only inside these fibres and zero elsewhere. Equivalently, in the Voigt matrix notation the total energy of the periodic unit cell reads
\begin{align}
	\mathcal{E} & = \frac{1}{2} \int_V \sum_{i=1}^{n} (\column{\varepsilon}^f_i - {}^h\column{\varepsilon}^f_i)^\mathsf{T} \, \mtrx{C}^f_i \, (\column{\varepsilon}^f - {}^h\column{\varepsilon}^f_i) \, \mathrm{d}V + \frac{1}{2} \int_A \sum_{j=1}^{s} \Delta\column{g}^\mathsf{T} \, \frac{k_h}{t} \, \Delta\column{g} \, \mathrm{d}A.
	\label{coheq4}
\end{align}

\begin{figure}
	\centering
	\subfloat[random fibre network]{\includegraphics[height=45mm]{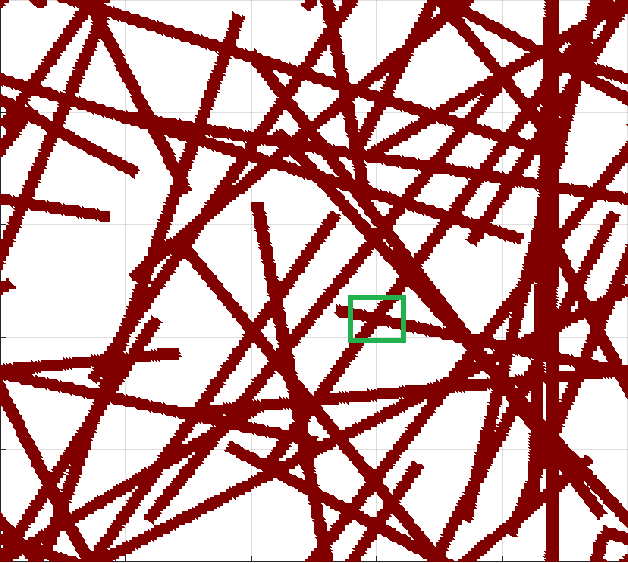}}
	\hspace{1em}
	\subfloat[zoomed-in bond region]{\includegraphics[height=45mm]{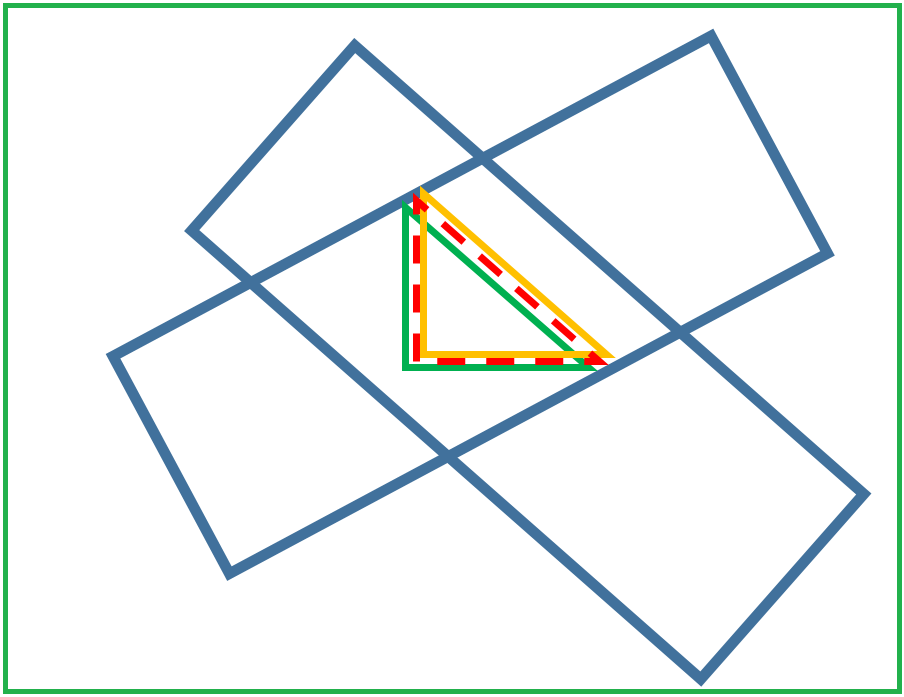}}
	\hspace{1em}
	\subfloat[an interfacial element]{\includegraphics[height=45mm]{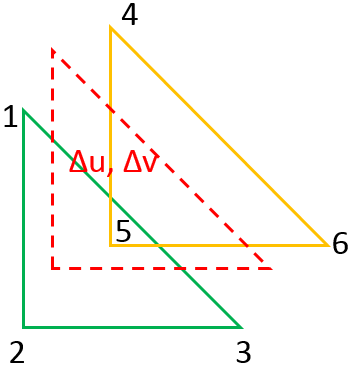}}
	\caption{(a)~A randomly generated fibrous paper network with one of the bond regions highlighted by the green rectangle. (b)~Zoom into the bond region of two fibres from~(a). (c)~A single interfacial element considered inside the bond region; top-fibre element is depicted in green, bottom-fibre element in yellow, while the interfacial element bonding the two fibres is depicted in dashed red.}
	\label{inte4}
\end{figure}

In Fig.~\ref{inte4}, part of the bonded region of two fibres in the network is considered, in which two triangular finite elements are depicted. These are connected through an interfacial element at their nodes. The components of the relative displacement~$\Delta \vec{g}$ for an interfacial element can be represented by
\begin{align}
	\Delta u & = \xi(u_1-u_4)\ + \ \eta(u_2-u_5)\ + \ (1-\xi-\eta)(u_3-u_6), \label{eq:deltau} \\
	\Delta v & = \xi(v_1-v_4)\ + \ \eta(v_2-v_5)\ + \ (1-\xi-\eta)(v_3-v_6), \label{eq:deltav}
\end{align}
where~$\xi$ and~$\eta$ are the local coordinates associated with the interfacial element. The relations~\eqref{eq:deltau} and~\eqref{eq:deltav} can be represented in matrix form as
\begin{equation}
	\Delta\underline{g} =
	\begin{pmatrix} \Delta u \\ \Delta v \end{pmatrix}
	= \mtrx{B}^c\column{w},
\end{equation}
where
\begin{multline}
	\mtrx{B}^c =
	\left(\begin{array}{cccccc}
			\xi & 0   & \eta & 0    & (1-\xi-\eta) & 0            \\
			0   & \xi & 0    & \eta & 0            & (1-\xi-\eta)
		\end{array}\right.
	\\
	\left.\begin{array}{cccccc}
			-\xi & 0    & -\eta & 0     & -(1-\xi-\eta) & 0             \\
			0    & -\xi & 0     & -\eta & 0             & -(1-\xi-\eta)
		\end{array}\right),
	\label{matrix44}
\end{multline}
and the nodal displacements at the nodes of both the connected finite elements are represented by the column matrix $\column{w} = [ u_1, v_1, u_2, v_2, u_3, v_3, u_4, v_4, u_5, v_5, u_6, v_6 ]^\mathsf{T}$. Upon substitution of~$\Delta\column{g}$ into the second term of Eq.~\eqref{coheq4}, the expression for the energy of all interfacial elements over all bonds is obtained.

Discretizing the entire network domain with~$m$ finite elements, $s$ interfacial elements, and taking the derivative of the elastic energy~$\mathcal{E}$ in Eq.~\eqref{coheq4} with respect to all kinematic degrees of freedom of the entire network collected in a column matrix~$\column{u}$ in order to minimize it, the following system of linear equations results
\begin{equation}
	\frac{1}{2} \sum^m_{e=1} \sum^n_{i=1} \int \mtrx{B}^\mathsf{T}_e \mtrx{C}_i^f \mtrx{B}_e t \, \mathrm{d}A \, \column{u}
	+ \frac{1}{2} \sum^s_{j=1} \int (\mtrx{B}^c_{j})^\mathsf{T} \, \frac{k_h}{t} \, \mtrx{B}^{c}_{j} \, \mathrm{d}A \, \column{u}
	= \frac{1}{2} \sum^m_{e=1} \sum^n_{i=1} \int \mtrx{B}^\mathsf{T}_e \mtrx{C}_i^f \, {}^h\column{\varepsilon}^f_i t \, \mathrm{d}A,
	\label{glob44}
\end{equation}
or
\begin{equation}
	( \mtrx{K} + \mtrx{K}_c ) \, \column{u} = \column{F}_h.
	\label{ew4}
\end{equation}
In Eqs.~\eqref{glob44} and~\eqref{ew4}, $t$ denotes the thickness of the fibres, $\mtrx{K}$ is the stiffness matrix of the finite elements representing the fibres, $\mtrx{K}_c$ is the stiffness matrix of the~$s$ interfacial elements in the bonded regions, and~$\column{F}_h$ is the hygroscopic load column. The stiffness computation for each of the interfacial element is carried out by the six-point Gauss--Legendre quadrature rule for triangular finite elements. Finally, the linear system of Eq.~\eqref{ew4} is solved to determine the response of the network to changes in moisture content. The load column vector~$\column{F}_h$ can be also combined or replaced with a load vector pertaining to any external loads applied to the network. This is illustrated in the results Section~\ref{sec:tensileLoading}, where the effective stiffness response of the network is evaluated under tensile loading.

One of the main motivations for using interfacial elements is that the~$\mtrx{K}_c$ matrix changes with the mesh in such a way that the stiffness enforced in the bonded region remains the same for a fixed value of~$k_h$. Had the interfaces been modelled by individual springs, the number of springs would change with the discretization and the individual spring stiffness would have to be adapted to ensure the same stiffness in the bonded regions for different mesh refinements.
%
%
\section{Results and discussion}
\label{sec:results}
In this section, the influence of inter-fibre bonding on the macroscopic properties of the network is assessed through numerical results obtained using the bond model proposed in Section~\ref{sec:constitutive}. Initially, a simplified case of a two-fibre network, considered also in~\citep{Bosco1}, is analysed to illustrate the behaviour of the bond model at a local level in Section~\ref{sec:simplified}. Thereafter, complex networks of different coverages subjected to hygroscopic strains and tensile loading are analysed to evaluate the relation between the sheet level network response and the fibre bonding in Sections~\ref{sec:hygroExpansion}--\ref{sec:anisotropy}. Note that, in accordance with the model formulation of Section~\ref{sec:constitutive}, all results presented herein correspond to elastic states of the considered microstructures only, with no dissipation involved.
%
%
\subsection{Simplified two-fibre network}
\label{sec:simplified}
\begin{figure}
	\subfloat[discretized two-fibre unit cell]{\includegraphics[height=60mm]{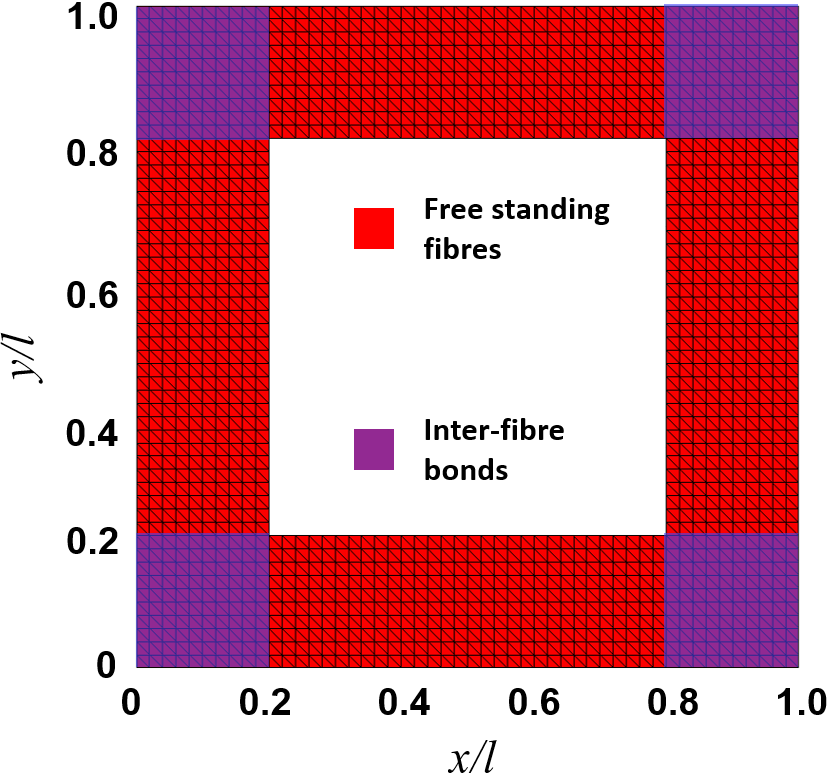}\label{fig14}}
	\hspace{1em}
	\subfloat[normalized local strains of individual fibres]{\includegraphics[height=60mm]{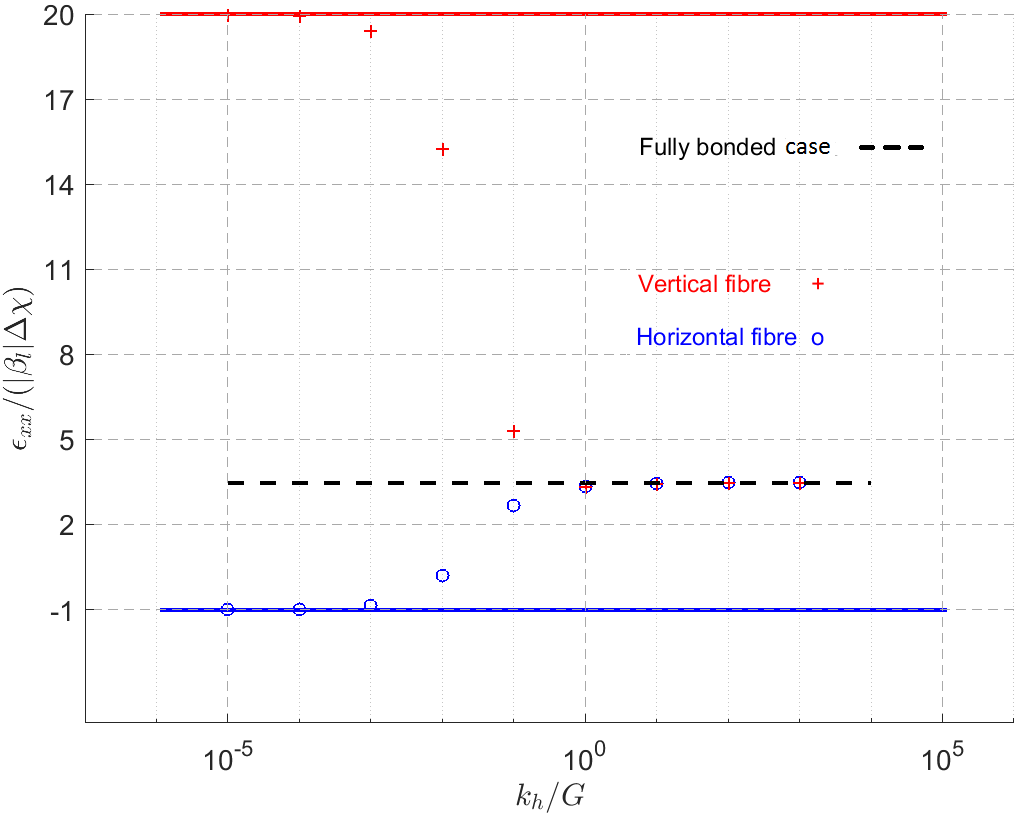}\label{fig24}}
	\caption{A two-fibre idealized network subjected to a change in moisture content~$\Delta\chi$. (a)~A discretized unit cell with~$51$ nodes in each direction. (b)~Normalized local strains in bonded region at ($ x/l = 1 $, $ y/l = 0 $) in the unit cell diagram (a) with the varying interfacial element stiffness modulus~$k_h$.}
\end{figure}

To analyse the effective hygro-expansivity, a two-fibre periodic cell is considered, i.e., the simplified mesoscale network of paper used earlier by~\citep{Bosco1}. Due to periodicity, the unit cell contains only half-fibres near the edges of the RVE, shown in Fig.~\ref{fig14}. The main reason to use this case is the simple nature of the mesoscale network, which helps to understand easily the effect of the bond stiffness on the local strain distribution of the fibres in the bonded regions. The mechanical behaviour of the fibres is assumed transversely isotropic with the elastic modulus in the longitudinal direction~$E_l$, elastic modulus in the transverse direction~$E_t = E_l/6 $ \citep{Strombo}, shear modulus~$ G_{lt} = E_l/10 $, Poisson's ratios~$ \nu_{lt} =0.3 $ and~$ \nu_{tl} = 0.05 $ \citep{Schulgasser}, and a thickness~$ t = l_f/50 $. The coefficient of hygro-expansion adopted for the fibres is also anisotropic with~$ \beta_l = -0.03 $ (based on the~\citealt{Jentzen} creep tests on single fibres; see also~\citealt{Sam3}), and~$ \beta_t = 20 \, |\beta_l| = 0.6 $. The network is subjected to a uniform change in moisture content for different values of the bond interface element modulus~$k_h$ in the inter-fibre bonds. The periodic unit cell is fixed in both directions at the left bottom corner, and fixed in the vertical direction at the right bottom corner, thus allowing free expansion in both the horizontal and vertical directions of the entire RVE, while removing all rigid body motions.

In the bond region located at the bottom and top right corner of the unit cell in Fig.~\ref{fig14}, the horizontal fibre has a low, and opposite, hygro-expansivity in the horizontal direction as compared to the vertical fibre. When the fibres are loosely bonded with a low value of~$k_h$, they behave independently. This leads to the development of negative local strains in the horizontal fibre, corresponding to~$-1$ when normalized with~$ |\beta_l| \, \Delta \chi $ as Fig.~\ref{fig24} shows. Similarly, a positive local strain (normalized) with a higher value corresponding to the ratio~$ | \beta_t/\beta_l | = 20 $ is reached in the horizontal direction for the vertical fibre.

As the bonding between the fibres becomes stiffer ($ k_h $ increases), an interaction occurs between the fibres due to their different directional expansion coefficients. This induces an increase in the local strains~$ \varepsilon_{xx} $ (normalized with~$ |\beta_l| \, \Delta \chi $) of the horizontal fibre and a decrease in the local strains~$ \varepsilon_{xx} $ in the vertical fibre. When the interfacial element modulus~$k_h$ is equal to the shear modulus of the fibres (representing more or less the 3D case with perfect bonding), the local strains in both fibres attain almost equal values. At even higher values of~$ k_h $, both local strains are exactly the same, attaining the value of strain identical to the one obtained for the same problem by assuming that the fibre mid-planes are rigidly bonded (as used by~\citealt{Sam2}, where no relative displacement was allowed). The obtained results are also qualitatively as well as quantitatively consistent with outcomes based on analytical expectations. It can be thus concluded that different bond properties~$k_h$ trigger significantly different local strains in individual fibres.
%
%
\subsection{Hygro-expansion of complex networks}
\label{sec:hygroExpansion}
\begin{figure}
	\centering
	\begin{tabular}{ll}
		\subfloat[coverage~$ c = 0.5 $]{\includegraphics[height=60mm]{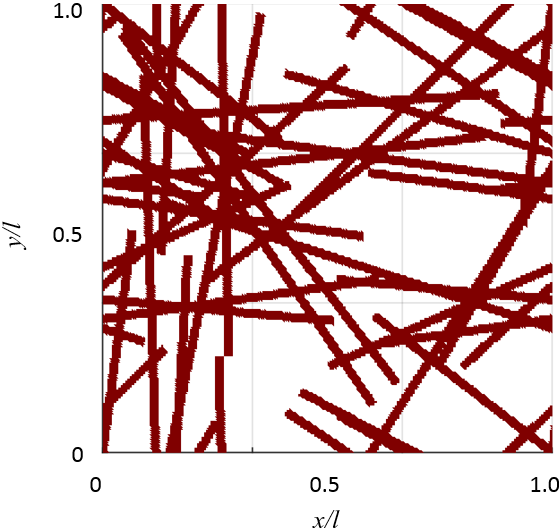}} &
		\subfloat[coverage~$ c = 1.0 $]{\includegraphics[height=60mm]{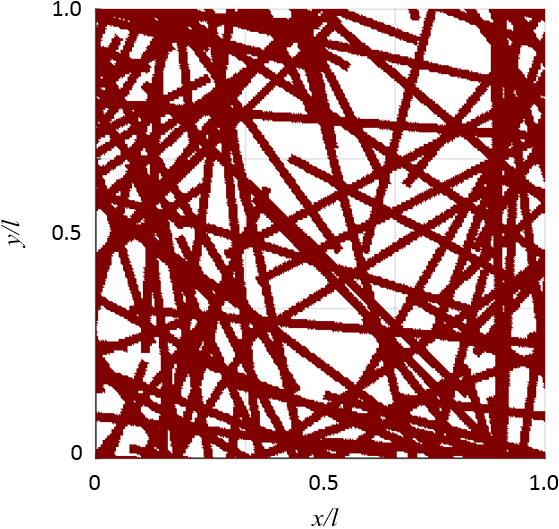}}    \\
		\subfloat[coverage~$ c = 2.0 $]{\includegraphics[height=60mm]{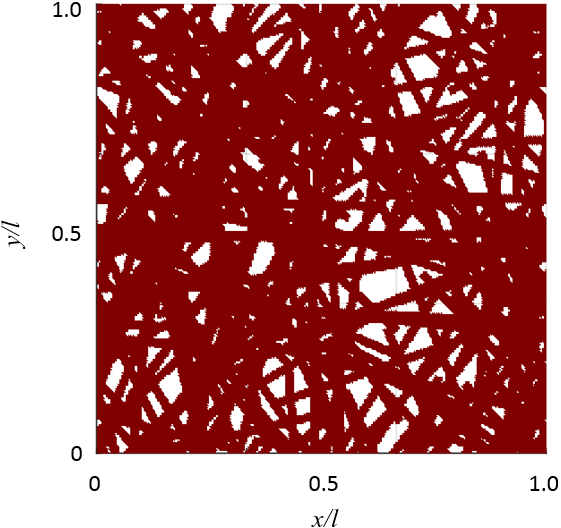}}  &
		\subfloat[coverage~$ c = 5.0 $]{\includegraphics[height=60mm]{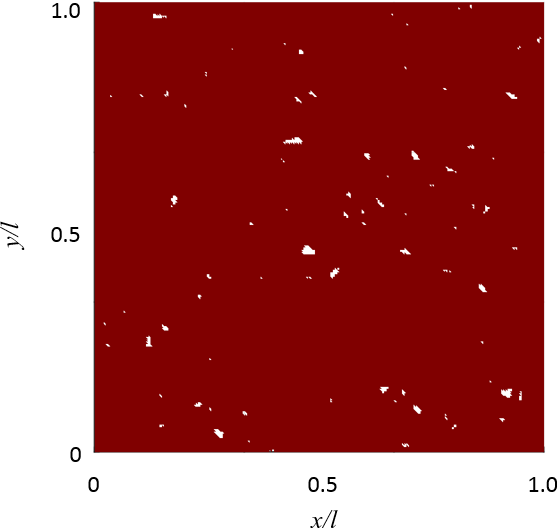}}
	\end{tabular}
	\caption{Random isotropic networks of different coverages~$ c = \{ 0.5, 1.0, 2.0, 5.0 \} $, considered for the representation of complex networks in a periodic~$ l \times l $ unit cell.}
	\label{ne44}
\end{figure}

More complex networks are considered next, with quasi isotropic orientation distributions of fibres and for coverages~$ c = \{ 0.5, 1.0, 2.0, 5.0 \} $, examples of which are shown in a periodic unit cell in Fig.~\ref{ne44}. Individual fibres are considered anisotropic with a longitudinal elastic modulus~$ E_l $, transverse elastic modulus~$ E_t = E_l/6 $, shear modulus~$ G = E_l/10 $, and the dimensions the same as mentioned earlier in Section~\ref{sec:networkModel}, i.e., $l = l_f$, $w_f = l_f/50$, and~$t = l_f/50$. The Poisson's ratios are again~$ \nu_{lt} = 0.3 $ and~$ \nu_{tl} = 0.05 $. The dimensions of the unit cell is~$ l \times l $. The finite element discretizations adopted for the complex networks were taken sufficiently fine to obtain converged results.

\begin{figure}
	\centering
	\includegraphics[height=70mm]{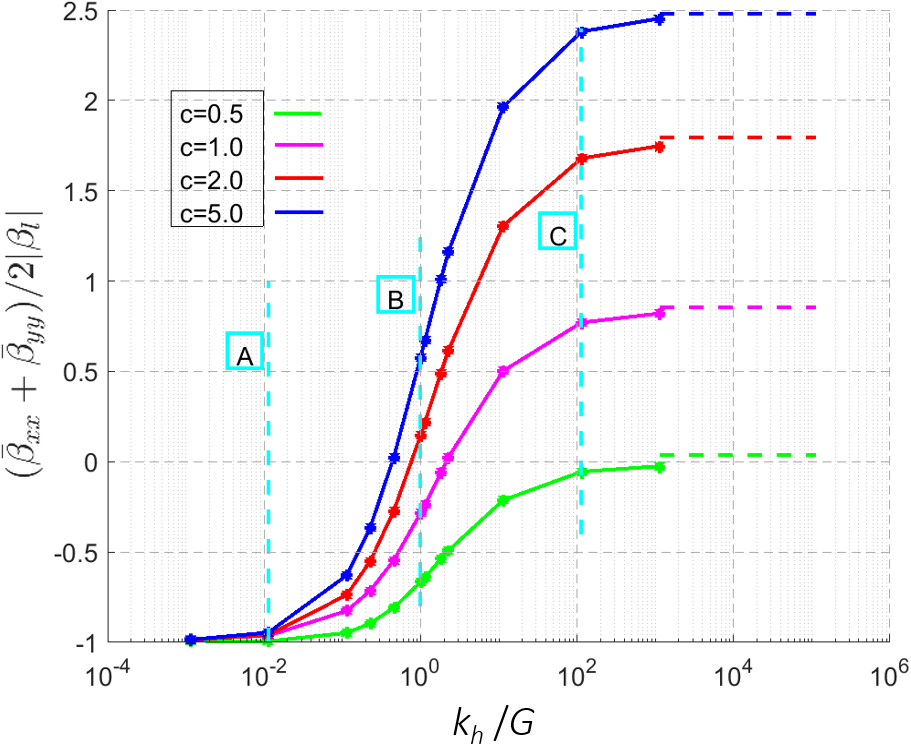}
	\caption{Normalized effective hygro-expansivity of the networks averaged in both directions versus the bond stiffness normalized by the shear modulus of fibres. The horizontal dashed lines represent the case of fully coupled bonding between individual fibres.}
	\label{hy4}
\end{figure}
\begin{figure}
	\centering
	\begin{tabular}{ll}
		\subfloat[region A, coverage~$ c = 1.0 $]{\includegraphics[height=60mm]{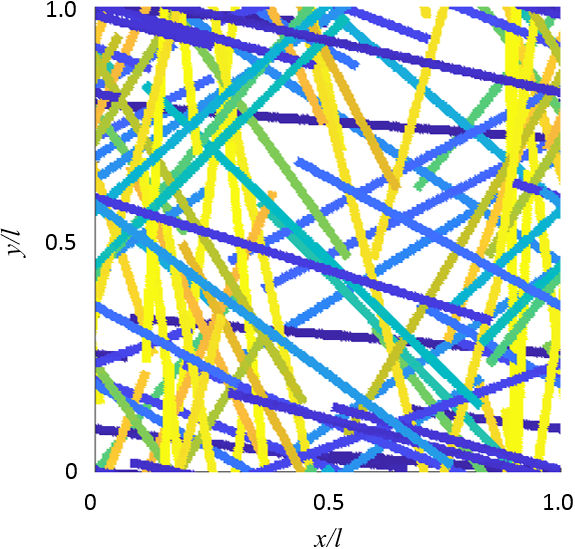}\label{oneone}}     &
		\subfloat[region B, coverage~$ c = 1.0 $]{\includegraphics[height=60mm]{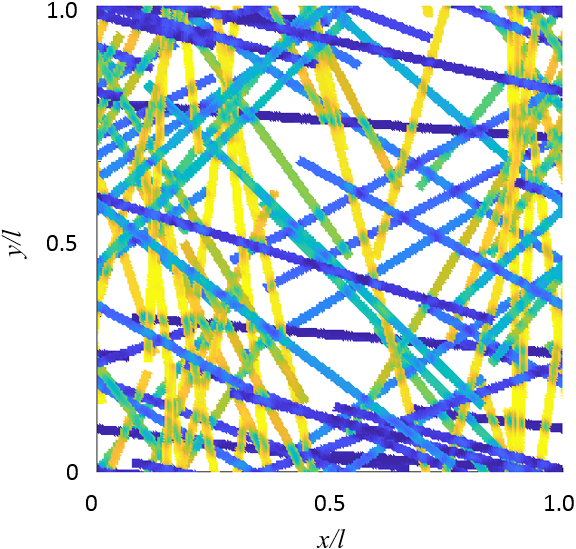}\label{twoone}}       \\
		\subfloat[region C, coverage~$ c = 1.0 $]{\includegraphics[height=60mm]{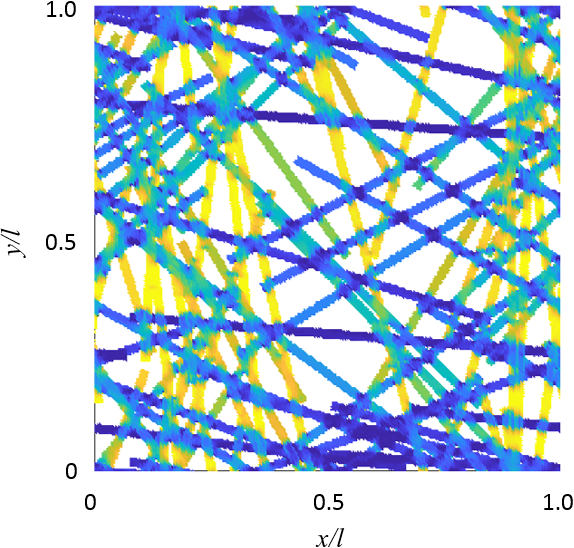}\label{threeone}} &
		\subfloat[fully bonded, coverage~$ c = 1.0 $]{\includegraphics[height=60mm]{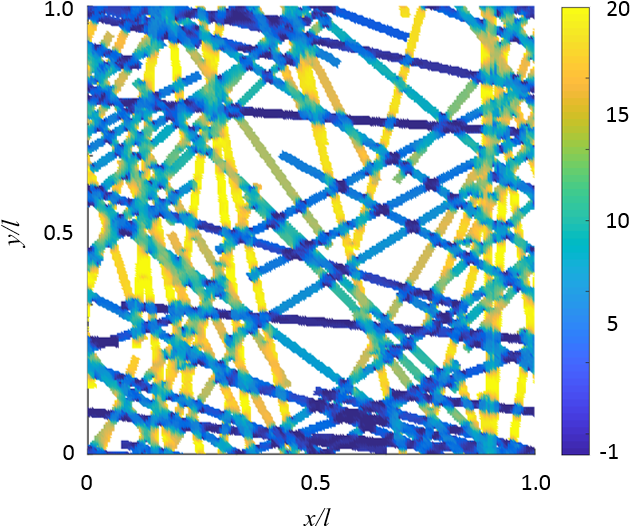}\label{fullone}}
	\end{tabular}
	\caption{Strain distribution~$\varepsilon_{xx} / |\beta_l|\Delta\chi$ in the network of coverage~$ c = 1.0 $ for individual regions A, B, and~C, and the fully bonded case, as indicated in Fig.~\ref{hy4}.}
	\label{str11}
\end{figure}
\begin{figure}
	\centering
	\begin{tabular}{ll}
		\subfloat[region A, coverage~$ c = 5.0 $]{\includegraphics[height=60mm]{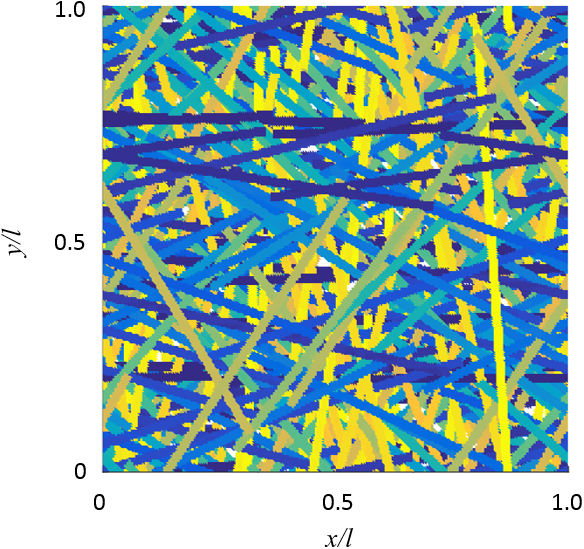}\label{onefive}}
		\subfloat[region B, coverage~$ c = 5.0 $]{\includegraphics[height=60mm]{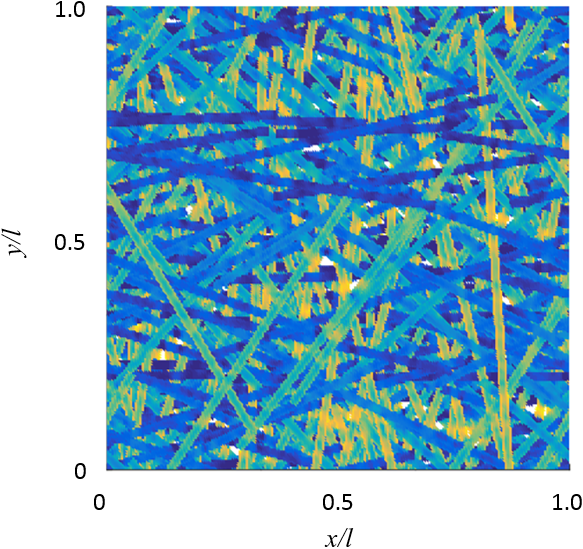}\label{twofive}} \\
		\subfloat[region C, coverage~$ c = 5.0 $]{\includegraphics[height=60mm]{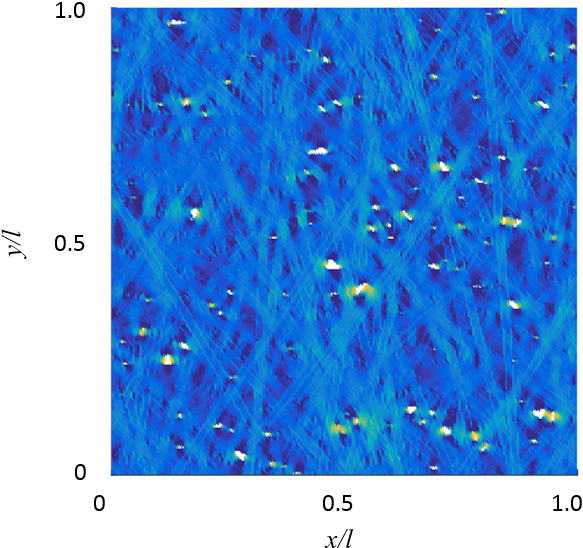}\label{threefive}}
		\subfloat[fully bonded, coverage~$ c = 5.0 $]{\includegraphics[height=60mm]{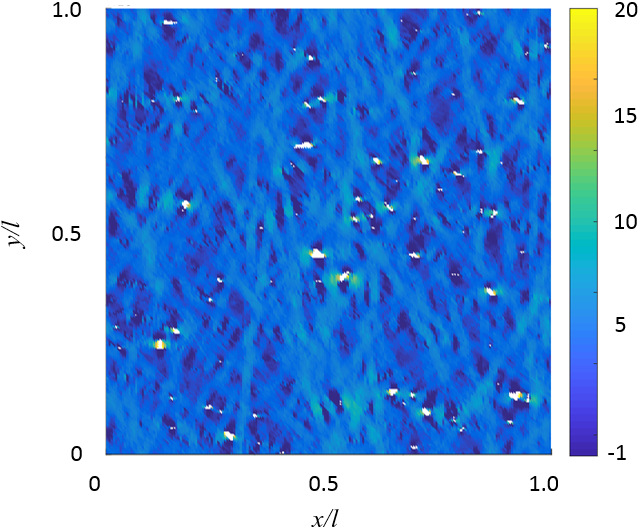}\label{fullfive}}
	\end{tabular}
	\caption{Strain distribution~$\varepsilon_{xx} / |\beta_l|\Delta\chi$ in the network of coverage~$ c = 5.0 $ for individual regions A, B, and~C, and the fully bonded case, as indicated in Fig.~\ref{hy4}.}
	\label{str14}
\end{figure}

The hygro-mechanical response of the complex networks with finite stiffness bonds is assessed by subjecting them to a uniform change in moisture content using the same boundary conditions as in the simplified two-fibre network, i.e., for free expansion. A range of values of~$k_h$ was used for the inter-fibre bonds, see Fig.~\ref{hy4}. For all coverages, at low values of the interfacial element modulus~$k_h$ (region A in Fig.~\ref{hy4}), the normalized effective hygro-expansivity in both directions is equal to the longitudinal swelling of a single fibre that is completely free in the considered periodic framework. At low~$k_h$, the fibres are not bonded or very loosely connected with each other. There is therefore no interaction between the transverse swelling of a given fibre and the longitudinal swelling of another fibre. As a result, the high transverse swelling of fibres cannot be transferred through the network and the network expands by longitudinal swelling of fibres only. This situation approximates fibre bonds as pin-joints only, where the longitudinal expansion governs the response of the entire network: the lateral expansion of fibres then does not have any influence on the macroscopic response of the network. This explains the lower hygro-expansivity in both directions in the network. Figures~\ref{oneone} and~\ref{onefive} furthermore show that fibres in the network have rather independent strains (normalized with~$ |\beta_l| \, \Delta \chi $), as a consequence of the absence of bonding between them.

With an increase of the interfacial modulus~$k_h$ to a value corresponding to the fibre shear modulus~$G$ (region B in Fig.~\ref{hy4}), an increased interaction between individual fibres takes place. As a result, the transverse hygroscopic swelling of one fibre may cause another fibre to have a positive strain in the longitudinal direction in these inter-fibre bonds. Therefore, a larger average hygro-expansivity of the networks emerges. In Figs.~\ref{twoone} and \ref{twofive}, the strain distributions in the bond regions and free standing parts of fibres of the network with $ c = 1.0 $ and $ c = 5.0 $ are affected by the partial kinematic constraint. It is worth noting that the network with higher coverages has a high fibre density, which causes a higher average expansivity as compared to lower coverages at this value of $ k_h $.

For a further increasing value of~$ k_h $, the average expansivity increases further for all coverages. The local strain distribution plots in Figs.~\ref{threeone} and~\ref{threefive} demonstrate more interaction between fibres, triggering different strain distributions in the fibre bond regions when compared to free standing regions. Note the tendency of fibres in a particular bond to attain identical strain values as the interfacial modulus is increased. As expected, the strain distribution in the network becomes similar to the case in Figs.~\ref{fullone} and~\ref{fullfive}, in which the fibres are rigidly bonded. Likewise, the effective hygro-expansivity tends towards the corresponding limit value in Fig.~\ref{hy4}. Hence, the sheet-scale results for rigid bonds are recovered as a limit case by the current bond model in the limit of $ k_h \to \infty $.
%
%
\subsection{Tensile loading of complex networks}
\label{sec:tensileLoading}
\begin{figure}
	\centering
	\includegraphics[height=70mm]{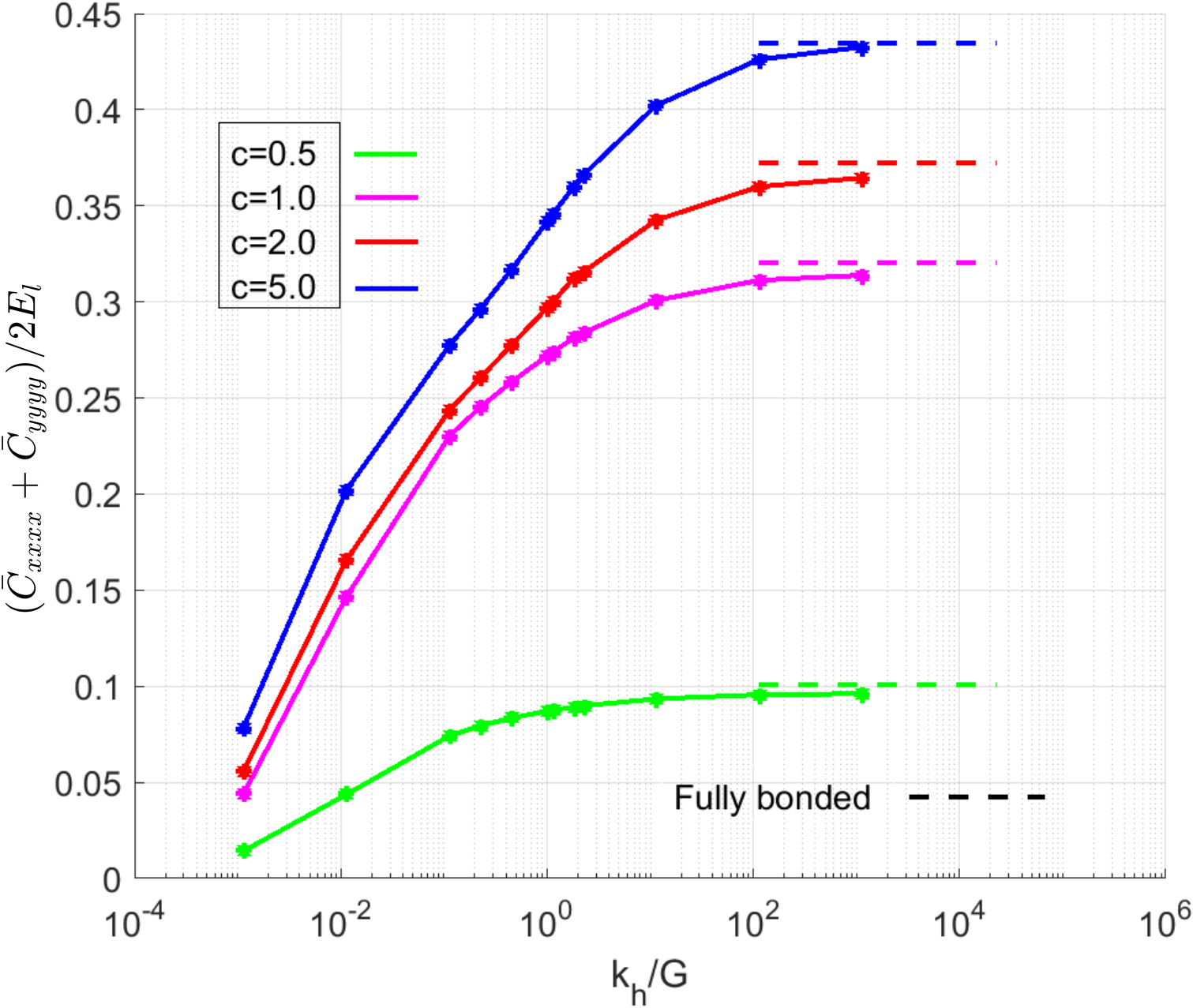}
	\caption{Normalized effective stiffness of networks averaged in both directions versus the bond stiffness normalized by the shear modulus of fibres. The horizontal dashed lines represent the case of fully coupled bonding between individual fibres.}
	\label{loa4}
\end{figure}

In order to understand the effect of the bonding stiffness between the fibres on the effective stiffness of the nearly isotropic networks, they are subjected to an external uniaxial tensile stress in the absence of hygroscopic strains. In the computation of the effective stiffness of the networks, the thickness taken into account is proportional to the coverage of the network, i.e., for~$ c = 5.0 $, the thickness is~$ 10 $ times as for coverage~$ c = 0.5 $. Two relevant stiffness constants, $ \overline{C}_{xxxx} $ and $ \overline{C}_{yyyy} $ of the effective elastic stiffness tensor ${}^4\overline{\tensor{C}}$ in Eq.~\eqref{constit4}, are computed from stress--strain diagrams of two distinct loading cases of the entire network (load applied in the $\vec{e}_x$ and $\vec{e}_y$ direction). As can be noticed in Fig.~\ref{loa4}, in the lower limit of the bond stiffness modulus $ k_h $, the fibres are loosely connected with each other at the inter-fibre bonds. This leads to less interaction between fibres, and hence a reduced ability to transfer stresses between fibres in the network. Therefore, both $ \overline{C}_{xxxx} $ and $ \overline{C}_{yyyy}  \to 0 $ as $ k_h \to 0 $. Since the effective stiffness of the fibrous network mostly depends on the ability of single and bonded fibres to transmit stresses, it results in a low effective stiffness of the networks for all coverages. As the bonding between the fibres becomes stiffer ($ k_h $ increases), the effective stiffness response of the networks also enhances, as expected. Like in the previous case, for high values of~$ k_h $, the average stiffness of the networks approaches the case of full kinematic constraint between fibres in the bonded regions. Furthermore, the higher coverages exhibit a higher stiffness due to the presence of more bonds as the number of fibres is higher. The observed findings on the dependence of the effective network stiffness on the bond stiffness~$k_h$ also qualitatively match with the results reported in~\citep[Fig.~7.11]{Heyden}, thus validating the model proposed herein. However, \cite{Heyden} used spring elements to accommodate finite bond stiffness instead of interfacial elements. Hence, for spring elements, the number of springs used and their stiffnesses need to be tuned appropriately, unlike interfacial elements that automatically adapt to different scales of discretization and bonding area.

Based on the response of the networks to hygro-expansion and tensile loading, several conclusions emerge. Firstly, a different effective response for each coverage results for different values of the bond stiffness. Therefore, even after relaxation of the kinematic constraints between fibres in the bonded regions of the network, the coverage still plays an essential role in determining the hygro-mechanical response of networks. This is consistent with earlier findings on the effect of coverage on hygro-expansivity \citep{Sam3,Sam2}. Next, for each coverage, the effective response of the networks at $ k_h = G $ (the value which describes the 3D case with perfect bonding) is substantially lower with respect to the 2D rigid case. This outlines the effect of the kinematics of the bond on the macroscopic response of the networks and highlights the limitation of the 2D rigid bond assumption.
%
%
\subsection{Anisotropy of complex networks}
\label{sec:anisotropy}
\begin{figure}
	\centering
	\includegraphics[height=40mm]{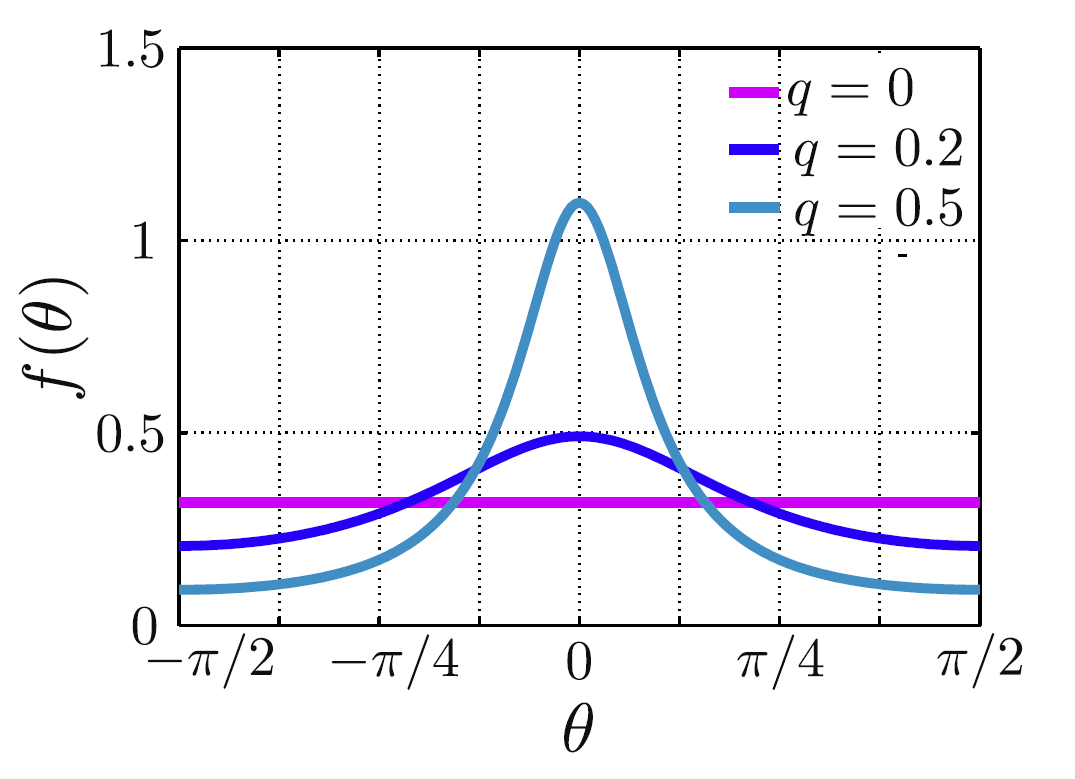}
	\caption{Probability density distribution function for fibre orientations in networks with different anisotropy levels~$q$, see Eq.~\eqref{eq:anisotropy} and~\citep{Bosco1}.}
	\label{ani4}
\end{figure}

The effect of the bond stiffness on the sheet scale response for anisotropic networks is finally studied. To this end, the orientation of the fibres in a network representing a paper sheet is described by a probability density function based on~\citep{Cox}, as shown in Fig.~\ref{ani4}, i.e.,
\begin{equation}
	f(\theta)=\frac{ 1-q^2 }{ \pi \big( 1+q^2-2q\cos(2\theta) \big) },
	\label{eq:anisotropy}
\end{equation}
where~$ \theta $ is the angle between the fibre and the machine direction, $ -\pi/2 < \theta < \pi/2 $, and~$ q $ is a measure of the anisotropy of the network.

\begin{figure}
	\centering
	\subfloat[difference in effective hygroexpansivity]{\includegraphics[height=65mm]{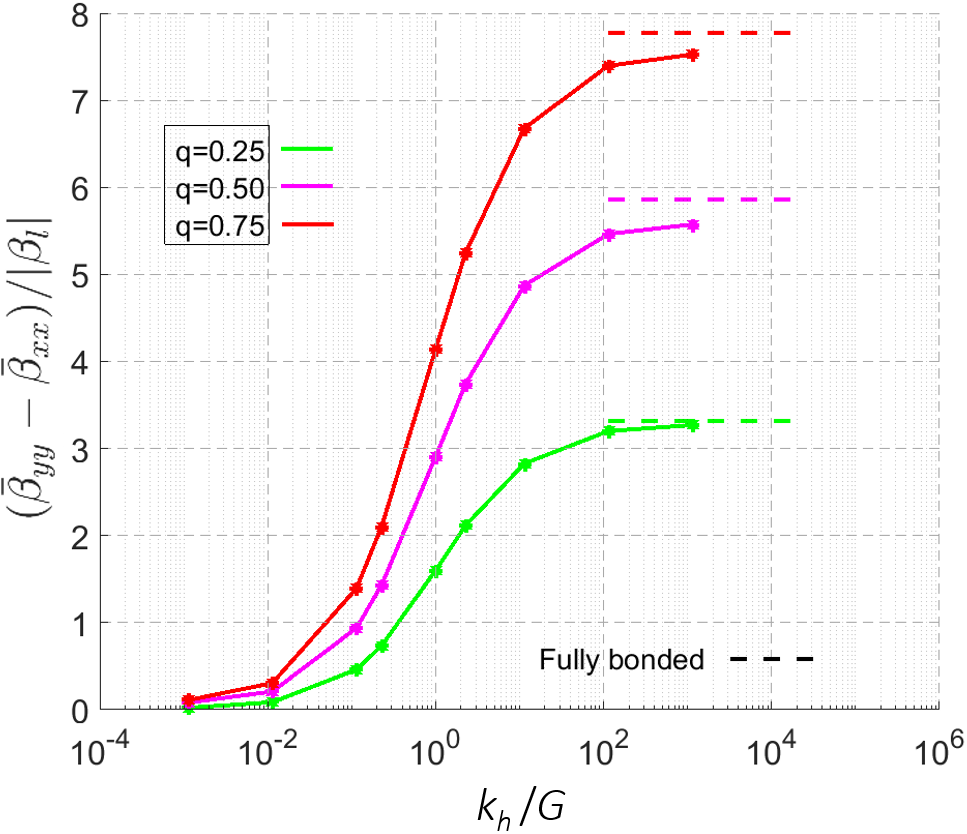}\label{aniso114}}
	\hspace{1em}
	\subfloat[difference in effective stiffness]{\includegraphics[height=65mm]{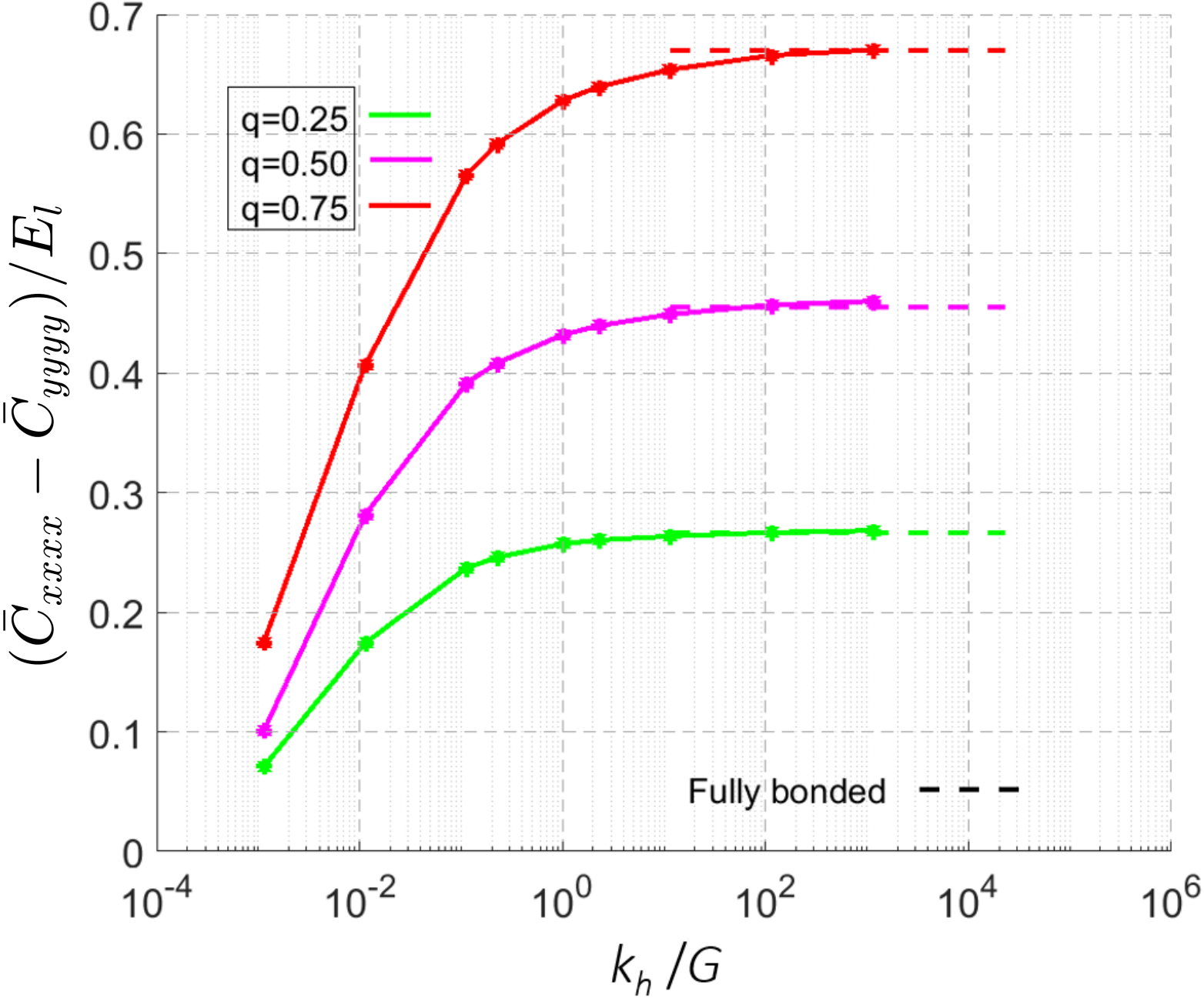}\label{aniso224}}
	\caption{Anisotropic response of networks with a varying interfacial element modulus~$ k_h $ normalized by the shear modulus of fibres~$G$. (a)~Difference of effective hygro-expansivity in both directions, normalized with~$ |\beta_l| $. (b)~Difference of effective stiffness in both directions, normalized with~$ E_l $.}
\end{figure}

Networks of coverage~$ c = 2.0 $ are considered, having three different degrees of anisotropy, $ q = \{ 0.25, 0.5, 0.75 \} $. These networks are subjected to a change in moisture content (hygroscopic loading) in one case and a uniaxial tensile loading in the other case, as done previously for various coverages. For rigidly connected inter-fibre bonds (depicted by the dashed horizontal lines in Fig.~\ref{aniso114}), more fibres oriented along the machine direction for $ q = 0.25 $ entail a larger expansion in the cross direction of the network at the sheet level, as $ \beta_t = 20 \, | \beta_l | $ for each individual fibre. The transverse expansivity of these fibres oriented close to the machine direction becomes more aligned with the network cross-direction for $ q = 0.25 $ as compared to $ q = 0.0 $. This effect becomes more pronounced for higher values of $ q $, leading to a higher hygro-expansivity in cross-direction of the networks for the case of full bonding.


At a low value of $ k_h $, the fibres in the networks behave like pin-jointed fibres exhibiting independent strains pertaining to the local longitudinal strain of each of the fibres ($ |\beta_l| \, \Delta\chi $, $ \Delta \chi = 1 $) in both the machine and cross directions irrespective of the values of $ q $. Therefore, in Fig.~\ref{aniso114}, it can be observed that the difference between the effective hygro-expansivity coefficients in both directions almost vanishes. As the bonding increases between fibres, the difference between the expansivity of the networks in the vertical and horizontal directions increases, implying a higher degree of anisotropy in the sheet response at higher~$ q $ values.

In terms of the effective stiffness of networks with different anisotropy subjected to tensile loading, the full kinematic constraint case for $ q = 0.25 $ (i.e., with more fibres oriented along the machine direction as compared to cross-direction) results in a higher stiffness in the machine direction, since $ E_l = 6 E_t $ for individual fibres. Fig.~\ref{aniso224} shows that as the anisotropy increases, the effective stiffness increases in machine direction, and hence also the difference in effective stiffness between both directions. A similar trend is again observed in the effective stiffness for different values of the bond stiffness for each of the values of $ q $ under consideration. The assumption of fully coupled bonding between fibres is therefore not appropriate for predicting the anisotropic response of a paper fibre network because it depends significantly on the extent of kinematic constraints or bonding between the fibres in the network.
%
%
\section{Conclusions}
\label{sec:conclusions}
In this paper, the influence of bonding between fibres at inter-fibre bonds on the sheet level behaviour of paper fibrous networks subjected to tensile loading and hygro-expansion was studied. The response of individual fibres was considered to be hygro-elastic for the sake of simplicity.

Initially, a random fibrous network is generated with rectangular-shaped fibres. These fibres are generated sequentially such that each fibre lies in a separate layer. Periodic boundary conditions are imposed for all the configurations considered. The entire network is discretized with triangular finite elements, numbered in the same sequence as the fibres. In the bond regions additional triangular interfacial elements are added between each couple of fibres in contact, allowing both fibres to be partially coupled with relative displacements between them. The extent to which both fibres in a bond undergo relative displacements depends on the stiffness modulus $ k_h $ of the bond represented by the interfacial elements. In order to investigate the influence of the stiffness modulus (i.e., bond stiffness) on the macroscopic response of the networks, numerical simulations for hygro-expansion and tensile uniaxial loading have been performed. The main conclusions can be summarized as follows:
\begin{enumerate}
	\item At low values of $ k_h $, the macro-level hygro-expansive response corresponds to the individual fibre longitudinal expansion in a free state. At high values of $ k_h $, the corresponding macro-level response tends towards the fully coupled case for both hygro-expansion and tensile load cases.

	\item Given the substantial difference between the effective hygro-expansivity of a particular network at $ k_h = G $ and the fully bonded case, it is concluded that the extent of kinematic constraints in the bonds is important in  influencing the sheet-scale behaviour of the network. Hence, the bond model with $ k_h = G $ (representing the physics between the mid-plane of fibres) is a more realistic approach for predicting the hygro-mechanical behaviour of paper.

	\item At different coverages, a different effective response of the network results for the bond model. This is in accordance with earlier findings \citep{Sam1,Sam2}.

	\item The anisotropic response at the sheet level also increases with an increase in $ k_h $ and vice-versa, highlighting significance of the bonding between fibres.
\end{enumerate}

Clear insights have been therefore obtained on the variability of the sheet scale hygro-mechanical response of fibrous networks at different values of the bonding stiffness between fibres. The degree to which the anisotropic response of the network is affected by the bond modulus in the inter-fibre bonds is also studied for some anisotropic orientation distributions of fibres. This study showed that, using the proposed bond model, an adequate representation of 3D fibres in a two dimensional framework can be obtained with relatively low computational efforts.

The presented work offers further scope of extensions by adopting a  moisture dependent interfacial element modulus to study its role on the effective hygro-expansivity of paper networks. Another possible scope of future work is to investigate hygro-inelastic response of the networks, which requires modelling of the plastic deformation in the fibres~\citep{Sam1}, still preserving the interfacial elements in bonds to understand their role in the development of irreversible deformations in the network. Finally, a non-linear response of the interfacial elements can be accounted for, involving a realistic damage model. All mentioned extensions aim at capturing the underlying physics of the fibre networks as accurately as possible with efficient two-dimensional models.
%
%
\section{Acknowledgements}
The first author would like to acknowledge the financial support granted by the European Commission~EACEA, grant reference \textnumero~2013-0043, as a part of the EM Joint Doctorate Simulation in Engineering and Entrepreneurship Development~(SEED), as well as the financial support granted by the Materials Innovation Institute~(M2i) and Canon Production Printing.
%
%
\bibliography{mybibfile}

\begin{thebibliography}{50}
\expandafter\ifx\csname natexlab\endcsname\relax\def\natexlab#1{#1}\fi
\providecommand{\url}[1]{\texttt{#1}}
\providecommand{\href}[2]{#2}
\providecommand{\path}[1]{#1}
\providecommand{\DOIprefix}{doi:}
\providecommand{\ArXivprefix}{arXiv:}
\providecommand{\URLprefix}{URL: }
\providecommand{\Pubmedprefix}{pmid:}
\providecommand{\doi}[1]{\href{http://dx.doi.org/#1}{\path{#1}}}
\providecommand{\Pubmed}[1]{\href{pmid:#1}{\path{#1}}}
\providecommand{\bibinfo}[2]{#2}
\ifx\xfnm\relax \def\xfnm[#1]{\unskip,\space#1}\fi
\bibitem[{Alince(2002)}]{Alince}
\bibinfo{author}{Alince, B.}, \bibinfo{year}{2002}.
\newblock \bibinfo{title}{Porosity of swollen pulp fibres revisited}.
\newblock \bibinfo{journal}{Nordic Pulp and Paper Research Journal}
  \bibinfo{volume}{12}, \bibinfo{pages}{3}.
\newblock \URLprefix
  \url{https://www.degruyter.com/document/doi/10.3183/npprj-2002-17-01-p071-073/html}.
\bibitem[{Bergstr{\"{o}}m(2018)}]{bergstrom}
\bibinfo{author}{Bergstr{\"{o}}m, P.}, \bibinfo{year}{2018}.
\newblock \bibinfo{title}{{Modelling mechanics of fibre network using discrete
  element method}}.
\newblock Ph.D. thesis. Mid Sweden University.
\bibitem[{Berthold et~al.(1994)Berthold, Desbrieres, Rinaudo and
  Salm\`en}]{Berthold}
\bibinfo{author}{Berthold, J.}, \bibinfo{author}{Desbrieres, J.},
  \bibinfo{author}{Rinaudo, M.}, \bibinfo{author}{Salm\`en, L.},
  \bibinfo{year}{1994}.
\newblock \bibinfo{title}{Types of adsorbed water in relation to the ionic
  groups and their counter-ions for some cellulose derivatives}.
\newblock \bibinfo{journal}{Polymer} \bibinfo{volume}{35}, \bibinfo{pages}{8}.
\newblock \URLprefix
  \url{https://www.sciencedirect.com/science/article/abs/pii/S0032386105800485}.
\bibitem[{Bosco et~al.(2015a)Bosco, Peerlings and Geers}]{Bosco3}
\bibinfo{author}{Bosco, E.}, \bibinfo{author}{Peerlings, R.H.J.},
  \bibinfo{author}{Geers, M.G.D.}, \bibinfo{year}{2015}a.
\newblock \bibinfo{title}{{Explaining irreversible hygroscopic strains in
  paper: a multi-scale modelling study on the role of fibre activation and
  micro-compressions}}.
\newblock \bibinfo{journal}{Mechanics of Materials} \bibinfo{volume}{91},
  \bibinfo{pages}{76--94}.
\newblock \URLprefix
  \url{https://linkinghub.elsevier.com/retrieve/pii/S016766361500157X},
  \DOIprefix\doi{10.1016/j.mechmat.2015.07.009}.
\bibitem[{Bosco et~al.(2015b)Bosco, Peerlings and Geers}]{Bosco1}
\bibinfo{author}{Bosco, E.}, \bibinfo{author}{Peerlings, R.H.J.},
  \bibinfo{author}{Geers, M.G.D.}, \bibinfo{year}{2015}b.
\newblock \bibinfo{title}{{Predicting hygro-elastic properties of paper sheets
  based on an idealized model of the underlying fibrous network}}.
\newblock \bibinfo{journal}{International Journal of Solids and Structures}
  \bibinfo{volume}{56-57}, \bibinfo{pages}{43--52}.
\newblock \URLprefix
  \url{https://linkinghub.elsevier.com/retrieve/pii/S0020768314004600},
  \DOIprefix\doi{10.1016/j.ijsolstr.2014.12.006}.
\bibitem[{Bosco et~al.(2017)Bosco, Peerlings and Geers}]{Bosco2}
\bibinfo{author}{Bosco, E.}, \bibinfo{author}{Peerlings, R.H.J.},
  \bibinfo{author}{Geers, M.G.D.}, \bibinfo{year}{2017}.
\newblock \bibinfo{title}{{Asymptotic homogenization of hygro-thermo-mechanical
  properties of fibrous networks}}.
\newblock \bibinfo{journal}{International Journal of Solids and Structures}
  \bibinfo{volume}{115-116}, \bibinfo{pages}{180--189}.
\newblock \URLprefix
  \url{https://linkinghub.elsevier.com/retrieve/pii/S0020768317301221},
  \DOIprefix\doi{10.1016/j.ijsolstr.2017.03.015}.
\bibitem[{Br{\"a}ndstr{\"o}m(2002)}]{Brandstorm}
\bibinfo{author}{Br{\"a}ndstr{\"o}m, J.}, \bibinfo{year}{2002}.
\newblock \bibinfo{title}{Morphology of norway spruce tracheids with emphasis
  on cell wall organisation}.
\newblock \bibinfo{journal}{Department of Wood Science, Acta Universitatis
  agriculture Sueciae, Silvestria} \bibinfo{volume}{237}, \bibinfo{pages}{39}.
\bibitem[{Bronkhorst(2003)}]{Bronkhorst}
\bibinfo{author}{Bronkhorst, C.A.}, \bibinfo{year}{2003}.
\newblock \bibinfo{title}{{Modelling paper as a two-dimensional
  elastic–plastic stochastic network}}.
\newblock \bibinfo{journal}{International Journal of Solids and Structures}
  \bibinfo{volume}{40}, \bibinfo{pages}{5441--5454}.
\newblock \URLprefix
  \url{https://linkinghub.elsevier.com/retrieve/pii/S0020768303002816},
  \DOIprefix\doi{10.1016/S0020-7683(03)00281-6}.
\bibitem[{Cave(1978)}]{Cave}
\bibinfo{author}{Cave, I.}, \bibinfo{year}{1978}.
\newblock \bibinfo{title}{Modelling moisture-related mechanical properties of
  wood part i: Properties of the wood constituents}.
\newblock \bibinfo{journal}{Wood Science and Technology} \bibinfo{volume}{12},
  \bibinfo{pages}{12}.
\newblock \URLprefix
  \url{https://link.springer.com/article/10.1007/BF00350818}.
\bibitem[{Chandrupatla and Belgundu(2002)}]{Chandru}
\bibinfo{author}{Chandrupatla, T.R.}, \bibinfo{author}{Belgundu, A.D.},
  \bibinfo{year}{2002}.
\newblock \bibinfo{title}{{Introduction to Finite Elements in Engineering}}.
\newblock \bibinfo{edition}{3rd} ed., \bibinfo{publisher}{Pearson}.
\bibitem[{Cox(1952)}]{Cox}
\bibinfo{author}{Cox, H.L.}, \bibinfo{year}{1952}.
\newblock \bibinfo{title}{{The elasticity and strength of paper and other
  fibrous materials}}.
\newblock \bibinfo{journal}{British Journal of Applied Physics}
  \bibinfo{volume}{3}, \bibinfo{pages}{72--79}.
\newblock \URLprefix
  \url{https://iopscience.iop.org/article/10.1088/0508-3443/3/3/302},
  \DOIprefix\doi{10.1088/0508-3443/3/3/302}.
\bibitem[{Eriksson et~al.(1991)Eriksson, Haglind, Lidbrandt and
  Sahn\`en}]{eriksson}
\bibinfo{author}{Eriksson, I.}, \bibinfo{author}{Haglind, I.},
  \bibinfo{author}{Lidbrandt, O.}, \bibinfo{author}{Sahn\`en, L.},
  \bibinfo{year}{1991}.
\newblock \bibinfo{title}{Fiber swelling favoured by lignin softening}.
\newblock \bibinfo{journal}{Wood Science and Technology} \bibinfo{volume}{25},
  \bibinfo{pages}{10}.
\newblock \URLprefix
  \url{https://agris.fao.org/agris-search/search.do?recordID=US19920055487}.
\bibitem[{Erkkila et~al.(2015)Erkkila, Leppänen, Ora, Tuovinen and
  Puurtinen}]{Erkilla}
\bibinfo{author}{Erkkila, A.}, \bibinfo{author}{Leppänen, T.},
  \bibinfo{author}{Ora, M.}, \bibinfo{author}{Tuovinen, T.T.},
  \bibinfo{author}{Puurtinen, A.}, \bibinfo{year}{2015}.
\newblock \bibinfo{title}{{Hygroexpansivity of anisotropic sheets}}.
\newblock \bibinfo{journal}{Nordic Pulp and Paper Research Journal}
  \bibinfo{volume}{30}, \bibinfo{pages}{326--334}.
\newblock \URLprefix \url{http://www.npprj.se/html/xml/toc10035.html},
  \DOIprefix\doi{10.3183/NPPRJ-2015-30-02-p326-334}.
\bibitem[{Fahl\`en and Salm\`en(2003)}]{Fahlen}
\bibinfo{author}{Fahl\`en, J.}, \bibinfo{author}{Salm\`en, L.},
  \bibinfo{year}{2003}.
\newblock \bibinfo{title}{Cross-sectional structure of the secondary wall of
  wood fibers as affected by processing.}
\newblock \bibinfo{journal}{Journal Of Materials Science} \bibinfo{volume}{38},
  \bibinfo{pages}{8}.
\newblock \URLprefix
  \url{https://link.springer.com/article/10.1023/A:1021174118468}.
\bibitem[{Gilani(2006)}]{Gilani}
\bibinfo{author}{Gilani, M.}, \bibinfo{year}{2006}.
\newblock \bibinfo{title}{A micromechanical approach to the behavior of
  singlewood fibers and wood fracture at cellular level.}
\newblock \bibinfo{journal}{Phd Thesis: \'{E}cole Polytechnique F\'{e}d\'{e}ral
  de Lausanne} \bibinfo{volume}{3546}, \bibinfo{pages}{168}.
\bibitem[{Glatfelter(2005)}]{bullet}
\bibinfo{author}{Glatfelter}, \bibinfo{year}{2005}.
\newblock \bibinfo{title}{Paper moisture and relative humidity}.
\newblock \URLprefix
  \url{https://pdf4pro.com/view/mwv-5460-tb2004-2-moisture-glatfelter-1f291b.html}.
\bibitem[{Hansen and Bj\"okman(1998)}]{hansen}
\bibinfo{author}{Hansen, C.}, \bibinfo{author}{Bj\"okman, A.},
  \bibinfo{year}{1998}.
\newblock \bibinfo{title}{The ultrastructure of wood from a solubility
  parameter point of view.}
\newblock \bibinfo{journal}{Wood Research And Technoology Holzforschung}
  \bibinfo{volume}{52}, \bibinfo{pages}{10}.
\newblock \URLprefix
  \url{https://www.degruyter.com/document/doi/10.1515/hfsg.1998.52.4.335/html}.
\bibitem[{Henriksson et~al.(2008)Henriksson, Berglund, Isaksson,
  Lindstr{\"{o}}m and Nishino}]{Henrikkson}
\bibinfo{author}{Henriksson, M.}, \bibinfo{author}{Berglund, L.A.},
  \bibinfo{author}{Isaksson, P.}, \bibinfo{author}{Lindstr{\"{o}}m, T.},
  \bibinfo{author}{Nishino, T.}, \bibinfo{year}{2008}.
\newblock \bibinfo{title}{{Cellulose Nanopaper Structures of High Toughness}}.
\newblock \bibinfo{journal}{Biomacromolecules} \bibinfo{volume}{9},
  \bibinfo{pages}{1579--1585}.
\newblock \URLprefix \url{https://pubs.acs.org/doi/10.1021/bm800038n},
  \DOIprefix\doi{10.1021/bm800038n}.
\bibitem[{Heyden(2000)}]{Heyden}
\bibinfo{author}{Heyden, S.}, \bibinfo{year}{2000}.
\newblock \bibinfo{title}{{Network modelling for the evaluation of mechanical
  properties of cellulose fibre fluff}}.
\newblock \bibinfo{type}{Phd thesis}. Lund University.
\bibitem[{Jentzen(1964)}]{Jentzen}
\bibinfo{author}{Jentzen, C.A.}, \bibinfo{year}{1964}.
\newblock \bibinfo{title}{{The Effect of Stress Applied During Drying on Some
  of the Properties of Individual Pulp Fibers}}.
\newblock \bibinfo{type}{Phd thesis}. Rensselaer Polytechnic Institute.
\bibitem[{Karako{\c{c}} et~al.(2017)Karako{\c{c}}, Hiltunen and
  Paltakari}]{Karakoc}
\bibinfo{author}{Karako{\c{c}}, A.}, \bibinfo{author}{Hiltunen, E.},
  \bibinfo{author}{Paltakari, J.}, \bibinfo{year}{2017}.
\newblock \bibinfo{title}{{Geometrical and spatial effects on fiber network
  connectivity}}.
\newblock \bibinfo{journal}{Composite Structures} \bibinfo{volume}{168},
  \bibinfo{pages}{335--344}.
\newblock \URLprefix
  \url{https://linkinghub.elsevier.com/retrieve/pii/S0263822316320578},
  \DOIprefix\doi{10.1016/j.compstruct.2017.02.062}.
\bibitem[{Koh and Oyen(2012)}]{Koh}
\bibinfo{author}{Koh, C.T.}, \bibinfo{author}{Oyen, M.L.},
  \bibinfo{year}{2012}.
\newblock \bibinfo{title}{{Branching toughens fibrous networks}}.
\newblock \bibinfo{journal}{Journal of the Mechanical Behavior of Biomedical
  Materials} \bibinfo{volume}{12}, \bibinfo{pages}{74--82}.
\newblock \URLprefix
  \url{https://linkinghub.elsevier.com/retrieve/pii/S1751616112000914},
  \DOIprefix\doi{10.1016/j.jmbbm.2012.03.011}.
\bibitem[{Kulachenko and Uesaka(2012)}]{Kulachenko}
\bibinfo{author}{Kulachenko, A.}, \bibinfo{author}{Uesaka, T.},
  \bibinfo{year}{2012}.
\newblock \bibinfo{title}{{Direct simulations of fiber network deformation and
  failure}}.
\newblock \bibinfo{journal}{Mechanics of Materials} \bibinfo{volume}{51},
  \bibinfo{pages}{1--14}.
\newblock \URLprefix
  \url{https://linkinghub.elsevier.com/retrieve/pii/S0167663612000683},
  \DOIprefix\doi{10.1016/j.mechmat.2012.03.010}.
\bibitem[{Larsson and W{\aa}gberg(2008)}]{Larsson}
\bibinfo{author}{Larsson, P.A.}, \bibinfo{author}{W{\aa}gberg, L.},
  \bibinfo{year}{2008}.
\newblock \bibinfo{title}{{Influence of fibre--fibre joint properties on the
  dimensional stability of paper}}.
\newblock \bibinfo{journal}{Cellulose} \bibinfo{volume}{15},
  \bibinfo{pages}{515--525}.
\newblock \URLprefix \url{http://link.springer.com/10.1007/s10570-008-9203-y},
  \DOIprefix\doi{10.1007/s10570-008-9203-y}.
\bibitem[{Lee and Jasiuk(2013)}]{Lee}
\bibinfo{author}{Lee, Y.}, \bibinfo{author}{Jasiuk, I.}, \bibinfo{year}{2013}.
\newblock \bibinfo{title}{{Apparent elastic properties of random fiber
  networks}}.
\newblock \bibinfo{journal}{Computational Materials Science}
  \bibinfo{volume}{79}, \bibinfo{pages}{715--723}.
\newblock \URLprefix
  \url{https://linkinghub.elsevier.com/retrieve/pii/S0927025613004382},
  \DOIprefix\doi{10.1016/j.commatsci.2013.07.037}.
\bibitem[{Lindner(2017)}]{Lindner}
\bibinfo{author}{Lindner, M.}, \bibinfo{year}{2017}.
\newblock \bibinfo{title}{Factors affecting the hygroexpansion of paper}.
\newblock \bibinfo{journal}{Journal Of Materials Science} \bibinfo{volume}{53},
  \bibinfo{pages}{26}.
\newblock \URLprefix
  \url{https://link.springer.com/content/pdf/10.1007/s10853-017-1358-1.pdf}.
\bibitem[{Liu et~al.(2011)Liu, Chen, Wang and Li}]{Liu}
\bibinfo{author}{Liu, J.X.}, \bibinfo{author}{Chen, Z.T.},
  \bibinfo{author}{Wang, H.}, \bibinfo{author}{Li, K.C.}, \bibinfo{year}{2011}.
\newblock \bibinfo{title}{{Elasto-plastic analysis of influences of bond
  deformability on the mechanical behavior of fiber networks}}.
\newblock \bibinfo{journal}{Theoretical and Applied Fracture Mechanics}
  \bibinfo{volume}{55}, \bibinfo{pages}{131--139}.
\newblock \URLprefix
  \url{https://linkinghub.elsevier.com/retrieve/pii/S0167844211000243},
  \DOIprefix\doi{10.1016/j.tafmec.2011.04.003}.
\bibitem[{Mao et~al.(2017)Mao, Goutianos, Tu, Meng, Chen and Peijs}]{Mao1}
\bibinfo{author}{Mao, R.}, \bibinfo{author}{Goutianos, S.},
  \bibinfo{author}{Tu, W.}, \bibinfo{author}{Meng, N.}, \bibinfo{author}{Chen,
  S.}, \bibinfo{author}{Peijs, T.}, \bibinfo{year}{2017}.
\newblock \bibinfo{title}{{Modelling the elastic properties of cellulose
  nanopaper}}.
\newblock \bibinfo{journal}{Materials \& Design} \bibinfo{volume}{126},
  \bibinfo{pages}{183--189}.
\newblock \URLprefix
  \url{https://linkinghub.elsevier.com/retrieve/pii/S0264127517304070},
  \DOIprefix\doi{10.1016/j.matdes.2017.04.050}.
\bibitem[{Marklund and Varna(2009)}]{Marklund}
\bibinfo{author}{Marklund, E.}, \bibinfo{author}{Varna, J.},
  \bibinfo{year}{2009}.
\newblock \bibinfo{title}{Modelling the hygroexpansion of aligned wood fibre
  composites}.
\newblock \bibinfo{journal}{Composite Science and Technology}
  \bibinfo{volume}{69}, \bibinfo{pages}{7}.
\newblock \URLprefix
  \url{https://www.sciencedirect.com/science/article/abs/pii/S0266353809000542}.
\bibitem[{Mihranyan et~al.(1978)Mihranyan, Llagostera, Karmhag, Str\"omme and
  Ek}]{Mihranyan}
\bibinfo{author}{Mihranyan, A.}, \bibinfo{author}{Llagostera, A.},
  \bibinfo{author}{Karmhag, R.}, \bibinfo{author}{Str\"omme, M.},
  \bibinfo{author}{Ek, R.}, \bibinfo{year}{1978}.
\newblock \bibinfo{title}{Modelling moisture-related mechanical properties of
  wood part i: Properties of the wood constituents}.
\newblock \bibinfo{journal}{Wood Science and Technology} \bibinfo{volume}{12},
  \bibinfo{pages}{12}.
\newblock \URLprefix
  \url{https://link.springer.com/article/10.1007/BF00350818}.
\bibitem[{Neagu and Gamstedt(2007)}]{Neagu}
\bibinfo{author}{Neagu, R.}, \bibinfo{author}{Gamstedt, E.},
  \bibinfo{year}{2007}.
\newblock \bibinfo{title}{Modelling of effects of ultrastructural morphology on
  the hygroelastic properties of wood fibres}.
\newblock \bibinfo{journal}{Journal Of Materials Science} \bibinfo{volume}{42},
  \bibinfo{pages}{21}.
\newblock \URLprefix
  \url{https://link.springer.com/article/10.1007/s10853-006-1199-9}.
\bibitem[{Niskanen(1998)}]{Niskanen1}
\bibinfo{author}{Niskanen, K.}, \bibinfo{year}{1998}.
\newblock \bibinfo{title}{{Paper Physics (Papermaking science and
  technology)}}.
\newblock \bibinfo{edition}{2nd} ed., \bibinfo{publisher}{Fapet Oy}.
\bibitem[{Nordman(1958)}]{Nordman}
\bibinfo{author}{Nordman, L.}, \bibinfo{year}{1958}.
\newblock \bibinfo{title}{{Laboratory investigations into the dimensional
  stability of paper}}.
\newblock \bibinfo{publisher}{Technical Association of the Pulp and Paper
  Industry}.
\bibitem[{Ostoja-Starzewski and Stahl(2000)}]{Starzewski}
\bibinfo{author}{Ostoja-Starzewski, M.}, \bibinfo{author}{Stahl, D.C.},
  \bibinfo{year}{2000}.
\newblock \bibinfo{title}{{Random fiber networks and special elastic orthotropy
  of paper}}.
\newblock \bibinfo{journal}{Journal of elasticity and the physical science of
  solids} \bibinfo{volume}{60}, \bibinfo{pages}{131--149}.
\newblock \DOIprefix\doi{https://doi.org/10.1023/A:1010844929730}.
\bibitem[{Perkins(1990)}]{Perkins}
\bibinfo{author}{Perkins, R.}, \bibinfo{year}{1990}.
\newblock \bibinfo{title}{Micromechanics models for predicting the elastic and
  strength behavior of paper materials}.
\newblock \bibinfo{journal}{MRS Online Proceedings Library}
  \bibinfo{volume}{197}, \bibinfo{pages}{20}.
\newblock \URLprefix
  \url{https://link.springer.com/article/10.1557/PROC-197-99}.
\bibitem[{Ramasubramanian and Perkins(1988)}]{Rama}
\bibinfo{author}{Ramasubramanian, M.}, \bibinfo{author}{Perkins, R.},
  \bibinfo{year}{1988}.
\newblock \bibinfo{title}{Computer simulation of the uniaxial elastic-plastic
  behavior of paper}.
\newblock \bibinfo{journal}{Journal of Engineering Materials and Technology}
  \bibinfo{volume}{110}, \bibinfo{pages}{7}.
\newblock \URLprefix
  \url{https://asmedigitalcollection.asme.org/materialstechnology/article-abstract/110/2/117/385099/Computer-Simulation-of-the-Uniaxial-Elastic?redirectedFrom=fulltext}.
\bibitem[{Roylance(1996)}]{Roylance}
\bibinfo{author}{Roylance, D.}, \bibinfo{year}{1996}.
\newblock \bibinfo{title}{{Mechanics of Materials}}.
\newblock \bibinfo{edition}{1st} ed., \bibinfo{publisher}{Wiley}.
\bibitem[{Salm{\'{e}}n et~al.(1987)Salm{\'{e}}n, Boman, Fellers and
  Htun}]{Salmen}
\bibinfo{author}{Salm{\'{e}}n, L.}, \bibinfo{author}{Boman, R.},
  \bibinfo{author}{Fellers, C.}, \bibinfo{author}{Htun, M.},
  \bibinfo{year}{1987}.
\newblock \bibinfo{title}{{The implications of fiber and sheet structure for
  the hygroexpansivity of paper}}.
\newblock \bibinfo{journal}{Nordic Pulp \& Paper Research Journal}
  \bibinfo{volume}{2}, \bibinfo{pages}{127--131}.
\newblock \URLprefix
  \url{https://www.degruyter.com/document/doi/10.3183/npprj-1987-02-04-p127-131/html},
  \DOIprefix\doi{10.3183/npprj-1987-02-04-p127-131}.
\bibitem[{Samantray et~al.(2021a)Samantray, Massart, Peerlings and
  Geers}]{Sam3}
\bibinfo{author}{Samantray, P.}, \bibinfo{author}{Massart, T.J.},
  \bibinfo{author}{Peerlings, R.H.J.}, \bibinfo{author}{Geers, M.G.D.},
  \bibinfo{year}{2021}a.
\newblock \bibinfo{title}{Modeling the effect of creep in paper fibres under
  the influence of external loading and changes in moisture}.
\newblock \bibinfo{journal}{Mechanics of Materials} \bibinfo{volume}{163},
  \bibinfo{pages}{104075}.
\newblock \DOIprefix\doi{10.1016/j.mechmat.2021.104075}.
\bibitem[{Samantray et~al.(2020)Samantray, Peerlings, Bosco, Geers, Massart and
  Roko{\v{s}}}]{Sam2}
\bibinfo{author}{Samantray, P.}, \bibinfo{author}{Peerlings, R.H.J.},
  \bibinfo{author}{Bosco, E.}, \bibinfo{author}{Geers, M.G.D.},
  \bibinfo{author}{Massart, T.J.}, \bibinfo{author}{Roko{\v{s}}, O.},
  \bibinfo{year}{2020}.
\newblock \bibinfo{title}{{Level Set-Based Extended Finite Element Modeling of
  the Response of Fibrous Networks Under Hygroscopic Swelling}}.
\newblock \bibinfo{journal}{Journal of Applied Mechanics} \bibinfo{volume}{87}.
\newblock \URLprefix
  \url{https://asmedigitalcollection.asme.org/appliedmechanics/article/doi/10.1115/1.4047573/1084678/Level-SetBased-Extended-Finite-Element-Modeling-of},
  \DOIprefix\doi{10.1115/1.4047573}.
\bibitem[{Samantray et~al.(2021b)Samantray, Peerlings, Massart and
  Geers}]{Sam1}
\bibinfo{author}{Samantray, P.}, \bibinfo{author}{Peerlings, R.H.J.},
  \bibinfo{author}{Massart, T.J.}, \bibinfo{author}{Geers, M.G.D.},
  \bibinfo{year}{2021}b.
\newblock \bibinfo{title}{{Micro-mechanical modeling of irreversible
  hygroscopic strain in paper sheets exposed to moisture cycles}}.
\newblock \bibinfo{journal}{International Journal of Solids and Structures}
  \URLprefix
  \url{https://linkinghub.elsevier.com/retrieve/pii/S0020768321001025},
  \DOIprefix\doi{10.1016/j.ijsolstr.2021.03.011}.
\bibitem[{Schulgasser and Page(1988)}]{Schulgasser}
\bibinfo{author}{Schulgasser, K.}, \bibinfo{author}{Page, D.H.},
  \bibinfo{year}{1988}.
\newblock \bibinfo{title}{{The influence of tranverse fibre properties on the
  in-plane elastic behaviour of paper}}.
\newblock \bibinfo{journal}{Composites Science and Technology}
  \bibinfo{volume}{32}, \bibinfo{pages}{279--292}.
\newblock \URLprefix
  \url{https://linkinghub.elsevier.com/retrieve/pii/0266353888900668},
  \DOIprefix\doi{10.1016/0266-3538(88)90066-8}.
\bibitem[{Stahl and Cramer(1998)}]{Stahl}
\bibinfo{author}{Stahl, D.}, \bibinfo{author}{Cramer, S.},
  \bibinfo{year}{1998}.
\newblock \bibinfo{title}{A three dimensional network model for a low density
  fibrous composite}.
\newblock \bibinfo{journal}{ASME - Journal of Engineering Materials and
  Technology} \bibinfo{volume}{120}, \bibinfo{pages}{5}.
\newblock \URLprefix
  \url{https://asmedigitalcollection.asme.org/materialstechnology/article-abstract/120/2/126/403185/A-Three-Dimensional-Network-Model-for-a-Low?redirectedFrom=fulltext}.
\bibitem[{Str{\"{o}}mbro and Gudmundson(2008)}]{Strombo}
\bibinfo{author}{Str{\"{o}}mbro, J.}, \bibinfo{author}{Gudmundson, P.},
  \bibinfo{year}{2008}.
\newblock \bibinfo{title}{{An anisotropic fibre-network model for
  mechano-sorptive creep in paper}}.
\newblock \bibinfo{journal}{International Journal of Solids and Structures}
  \bibinfo{volume}{45}, \bibinfo{pages}{5765--5787}.
\newblock \URLprefix
  \url{https://linkinghub.elsevier.com/retrieve/pii/S0020768308002576},
  \DOIprefix\doi{10.1016/j.ijsolstr.2008.06.010}.
\bibitem[{Thorpe et~al.(1976)Thorpe, Mark, Eusufzai and Perkins}]{mark}
\bibinfo{author}{Thorpe, J.}, \bibinfo{author}{Mark, R.},
  \bibinfo{author}{Eusufzai, A.}, \bibinfo{author}{Perkins, R.},
  \bibinfo{year}{1976}.
\newblock \bibinfo{title}{Mechanical properties of fiber bonds}.
\newblock \bibinfo{journal}{TAPPI} \bibinfo{volume}{59}, \bibinfo{pages}{5}.
\bibitem[{Uesaka(1994)}]{Uesaka1}
\bibinfo{author}{Uesaka, T.}, \bibinfo{year}{1994}.
\newblock \bibinfo{title}{{General formula for hygroexpansion of paper}}.
\newblock \bibinfo{journal}{Journal of Materials Science} \bibinfo{volume}{29},
  \bibinfo{pages}{2373--2377}.
\newblock \URLprefix \url{http://link.springer.com/10.1007/BF00363429},
  \DOIprefix\doi{10.1007/BF00363429}.
\bibitem[{Uesaka and Qi(1994)}]{Uesaka2}
\bibinfo{author}{Uesaka, T.}, \bibinfo{author}{Qi, D.}, \bibinfo{year}{1994}.
\newblock \bibinfo{title}{{Hygroexpansivity of paper -- Effects of
  fibre-to-fibre bonding}}.
\newblock \bibinfo{journal}{Journal of Pulp and Paper Science}
  \bibinfo{volume}{20}, \bibinfo{pages}{175--179}.
\bibitem[{Uetani and Yano(2011)}]{Zeta}
\bibinfo{author}{Uetani, K.}, \bibinfo{author}{Yano, H.}, \bibinfo{year}{2011}.
\newblock \bibinfo{title}{Zeta potential time dependence reveals the swelling
  synamics of wood cellulose nanofibrils}.
\newblock \bibinfo{journal}{Langmuir} \bibinfo{volume}{28},
  \bibinfo{pages}{10}.
\newblock \URLprefix \url{https://pubs.acs.org/doi/10.1021/la203404g}.
\bibitem[{Vainio and Paulapuro(2007)}]{Vainio}
\bibinfo{author}{Vainio, A.K.}, \bibinfo{author}{Paulapuro, H.},
  \bibinfo{year}{2007}.
\newblock \bibinfo{title}{{Interfiber bonding and fibre segment activation
  mechanism in paper}}.
\newblock \bibinfo{journal}{Nordic Pulp Paper Research Journal}
  \bibinfo{volume}{2}, \bibinfo{pages}{442--458}.
\bibitem[{Wang et~al.(2014)Wang, Liu and Lai}]{Ningling}
\bibinfo{author}{Wang, N.}, \bibinfo{author}{Liu, W.}, \bibinfo{author}{Lai,
  J.}, \bibinfo{year}{2014}.
\newblock \bibinfo{title}{An attempt to model the influence of gradual
  transition between cell wall layers on cell wall hygroelastic properties}.
\newblock \bibinfo{journal}{Journal Of Materials Science} \bibinfo{volume}{49},
  \bibinfo{pages}{10}.
\newblock \URLprefix
  \url{https://link.springer.com/article/10.1007/s10853-013-7885-5}.

\end{thebibliography}
\end{document}